\documentclass[%
 aip,
rsi,%
 amsmath,amssymb,
preprint,%
]{revtex4-1}

\usepackage{graphicx}
\usepackage{subfig}
\usepackage{float}
 \usepackage{color}
 \usepackage{bm}
\usepackage[mathlines]{lineno}

\begin{document}

\preprint{AIP/123-QED}

\title[Modeling shear instabilities and wave breaking using vortex method]{Nonlinear modeling of  wave-topography interactions,  shear instabilities and \\ shear induced wave breaking using vortex method}

\author{Divyanshu Bhardwaj}


\author{Anirban Guha}
 \email{anirbanguha.ubc@gmail.com}
\affiliation{
Environmental and Geophysical Fluids Group, Department of Mechanical Engineering, Indian Institute of Technology Kanpur, U.P. 208016, India.\\}%

\date{\today}

\begin{abstract}
Theoretical studies on linear shear instabilities often use simple  velocity and density profiles (e.g.\ constant, piecewise) for obtaining good qualitative and quantitative predictions of the initial disturbances. Furthermore, such simple profiles  provide a minimal model for obtaining a mechanistic understanding of otherwise elusive shear instabilities. However, except a few specific cases, the efficacy of simple profiles  has  remained limited to the linear stability paradigm. In this work we have proposed a general framework that can  simulate the fully nonlinear evolution of a variety of  stratified shear instabilities as well as wave-wave and wave-topography interaction problems having simple profiles. To this effect, we have modified the classical vortex method by extending the Birkhoff-Rott equation to multiple interfaces, and furthermore, have incorporated background shear across a density interface. The latter is more subtle, and originates from the understanding that Bernoulli's equation is  not just limited to irrotational flows, but can be modified to make it applicable for piecewise velocity profiles.    
We have solved diverse problems that can be essentially reduced to the multiple interacting interfaces paradigm, e.g.\ spilling and plunging breakers, stratified shear instabilities like Holmboe and Taylor-Caulfield, jet flows, and even wave-topography interaction problem like Bragg resonance. Free-slip boundary being  a vortex sheet, its effect can also be effectively captured using vortex method. We found that the minimal models capture key nonlinear features, e.g.\ wave breaking features like cusp formation and roll-ups, which are observed in  experiments and/or extensive simulations with smooth, realistic profiles. 
\end{abstract}


\maketitle

\section{Introduction}
Velocity shear, with or without density stratification, are ubiquitous in environmental and industrial flows. Examples of shear flows include  pycnocline region in lakes and oceans, Antarctic circumpolar current, Stratospheric polar night jet, density currents,  fuel-air mixing layer in internal combustion engines. Stratified shear flows usually exhibit region(s) of strong gradients in background velocity and density, and are often unstable, leading to various types of shear instabilities. When gradients are very sharp, it leads to the formation of interfaces. A density interface is marked by a sharp change in background density  (e.g.\ air-sea interface), whereas a vorticity interface 
is produced due to a sharp change in background vorticity. A crucial feature of interfaces is that it can support progressive waves. A density interface can support two oppositely propagating gravity waves (e.g.\ the waves we observe on the ocean surface), whereas a vorticity interface supports a vorticity wave.  Wave breaking is a hallmark nonlinear feature of shear instabilities, and is crucial from application point of view. While surface gravity wave breaking is pivotal in understanding air-sea coupling \cite[]{melville1996role}, internal gravity wave breaking at the pycnocline plays a major role in understanding ocean/lake mixing and the subsequent biogeochemical processes \cite[]{smyth2012ocean}. In the atmosphere, Rossby wave (vorticity waves in a rotating frame) breaking is key to the understanding of ozone mixing, and the fate of the Antarctic ozone hole \cite[]{guha2016modeling}.

Stratified shear instabilities being reasonably complicated,  it is often helpful to theoretically simplify the background flow into piecewise linear and/or constant profiles.  In the linear regime (i.e., when the wave-like perturbations are infinitesimally small), a fairly accurate qualitative and quantitative description of shear instabilities can be obtained from  simplified piecewise profiles \cite{drazin2004}. 
 A pertinent question to ask is - do piecewise profiles also provide reasonably accurate description in the nonlinear regime as well? Since nonlinearity brings about additional complexities, it is not unreasonable to expect that the simplistic nature of piecewise background profiles may fail to capture the essential nonlinear features of shear instabilities emanating from continuous profiles. Fortunately the agreement is reasonably good for homogeneous flows (density is constant). A numerical technique known as  contour dynamics\cite{pozrikidis1985nonlinear,dritschel1989contour,pullin1992contour,guha2012}, developed around forty years ago \cite{deem1978vortex}, captures the 
 nonlinear evolution of shear instabilities emanating from piecewise linear velocity profiles (vorticity constant in each layer). The results obtained from contour dynamics simulations match well with that of shear instabilities ensuing from continuous shear layers. This breakthrough technique has been used to study a variety of homogeneous shear flow problems, including but not limited to geostrophic turbulence \cite{scott2012}. 

The success of the applicability of contour dynamics is however limited to homogeneous flows. The primary objective of this paper is to devise a generalized framework such that  density stratified shear instabilities emanating from simple piecewise profiles can be extended to the nonlinear regime. In other words, the goal is to simulate the nonlinear evolution of density interface(s) sandwiched in a shear layer. It is important to note here that the numerical strategy behind the nonlinear evolution of stable or unstable density interfaces (in absence of background velocity shear) is already well established, and is efficiently and accurately implemented using vortex methods \cite[]{saffman1992vortex}. 
This technique capitalizes on the fact that density interfaces are basically vortex sheets, hence can be discretized into a series of point vortices. Point vortices interact with each other; each point vortex moves with the resultant velocity imposed on it by all other vortices at any given time. Thus the interface defined by the point vortices also evolves in time.
A basic assumption here is that the flow is inviscid, and is therefore suitable in mimicking high Reynolds number flows occurring in geophysical settings.
The computation is also numerically efficient since the equations of motion are solved only at the interfaces, thereby reducing the dimensionality of the problem by one. In the past, vortex methods have been successfully implemented to accurately simulate the dynamics of a single unstable interface, e.g.\ Rayleigh-Taylor instability \cite[]{baker1980vortex,tryggvason1988numerical}, Richtymer-Meshkov instability \cite[]{sohn2004vortex}, Kelvin-Helmholtz instability (KH) \cite[]{sohn2010long}, as well as that of a stable interface, e.g.\ propagation of surface gravity waves \cite[]{baker1982generalized}. 

Before proceeding further we briefly summarize the advantages and shortcomings of the two above-mentioned techniques, namely contour dynamics and vortex method. 
While contour dynamics allows multiple vorticity interfaces, it is only applicable to homogeneous flows.
Vortex method, however, can capture the evolution of a single density interface (hence can model stratified  flows) provided there is no \emph{finite} background shear. The method can only capture infinite shear (jump in background velocity)  with or without density jump, for example, the classic KH instability set-up \cite{sohn2010long}. 
This is a major drawback, especially for simulating stratified shear instabilities, since in such flows, one or more density interfaces are usually embedded in a  shear layer. The most general scenario of piecewise density stratified shear layer is shown in figure \ref{fig:schematic_full}, which even includes a bottom topography. The interface at $z=\eta_2$ is the infinite shear interface, and as already mentioned, only the evolution of this interface can be captured using vortex method.  The fact that  restricts vortex method to shearless (or irrotational) background flow is the unsteady Bernoulli's equation, which governs the evolution of a vortex sheet.  The fundamental assumption behind the derivation of Bernoulli's equation is that \emph{ the flow must be irrotational}. 
Hence our aim in this paper is threefold : 
\\(i) To find the evolution equation of a vortex sheet in presence of shear. In other words, we need to generalize the well known unsteady Bernoulli's equation, which is \emph{only} applicable to irrotational background velocity profile, to profiles that are piecewise linear. 
\\ (ii) To find the interaction between multiple vortex sheets. This is because shear instabilities arise from the interaction between multiple interfaces in presence of  background shear (rather, waves present at these interfaces)\cite{caul1994,bain1994,heif2005,guha2014}. 
A point worth mentioning is that free-slip boundary is also a vortex sheet\cite{baker1982generalized}, hence an algorithm that captures the interaction between multiple vortex sheets would automatically include the interaction between a free interface and a solid boundary.
\\\ (iii) Once the above two conditions are fulfilled, it remains to be shown whether interacting density interfaces embedded in a shear layer replicate the key nonlinear features of shear instabilities, at least qualitatively.

The organization of the paper is as follows. In  section \ref{sec:Math_form} we  discuss the  mathematical formulation and the numerical technique. To show the versatility of the technique in capturing wave nonlinearities, including breaking under very different scenarios, we  broadly discuss three kinds of settings - flows with finite shear (sections \ref{sec:finite_shear} -   \ref{sec:finite_shear_internal}), strongly sheared flows (section \ref{sec:infinite_shear}), and flows without shear (section \ref{sec:No_shear}). In  section \ref{sec:finite_shear} we  capture surface wave breaking due to shear for both short waves (in the presence of surface tension) as well as long waves (where bottom topography becomes crucial). These are respectively known as the spilling and plunging breakers. In  section \ref{sec:finite_shear_internal} we  analyze the nonlinear evolution of two well known types of stratified shear instabilities that occur inside lakes and oceans - Holmboe instability and Taylor-Caulfield instability. In section \ref{sec:infinite_shear} we  discuss liquid sheets or jet flows that is susceptible to KH instability. Unlike previous studies, here KH instability occurs in more than one interfaces. In section \ref{sec:No_shear} we  highlight the importance of bottom topography by switching off background velocity. Here we  show a kind of wave-triad interactions, known as Bragg resonance, occurring between two oppositely propagating surface waves and the rippled bottom topography. The paper is  summarized and concluded in  section \ref{sec:summary}.

\section{Mathematical formulation and numerical procedure}
\label{sec:Math_form}

\begin{figure}
\centering
\includegraphics[width=140mm]{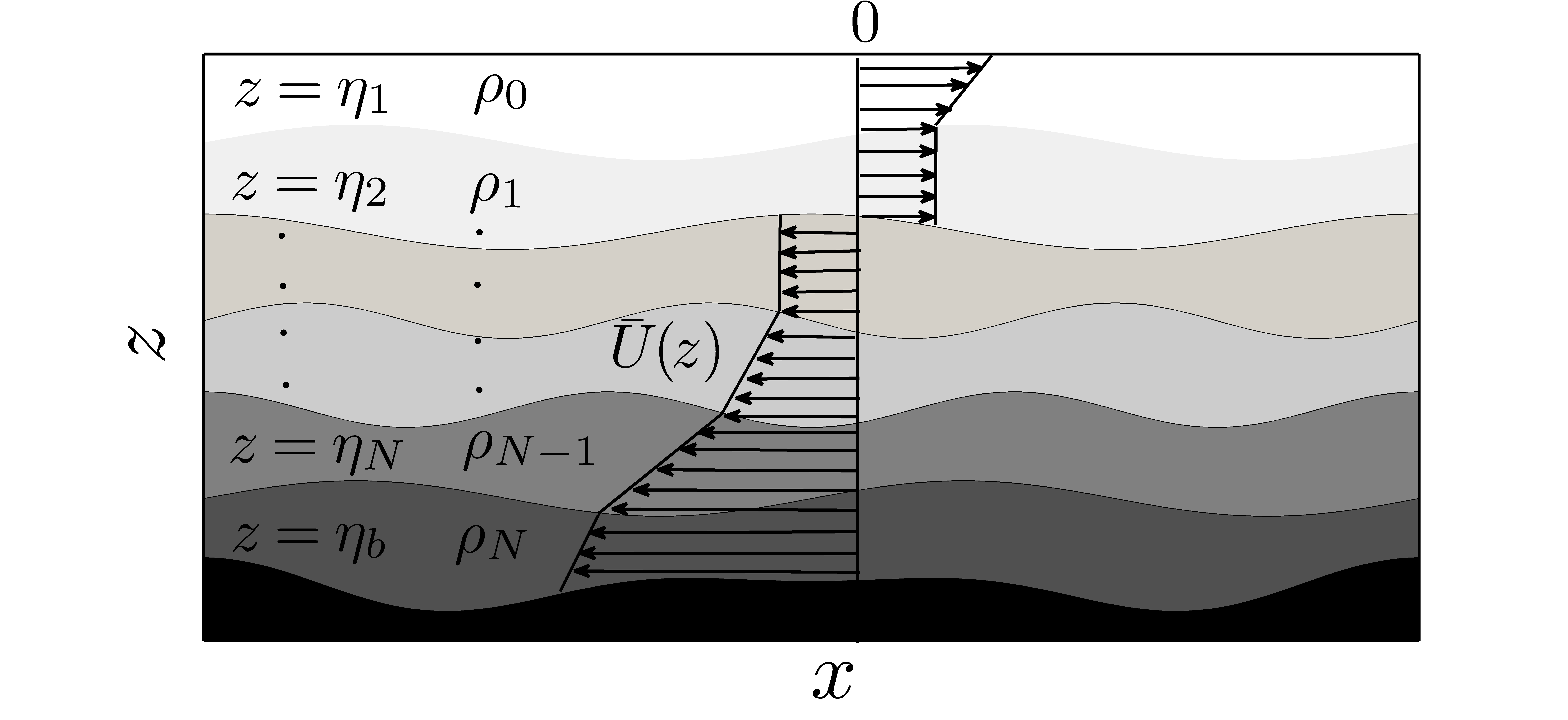}
\caption{A general schematic of our ``layered system'' having a piecewise constant density and piecewise linear base velocity profile. The bottommost interface $z=\eta_b$ denotes topography.}
\label{fig:schematic_full}
\end{figure}

We consider an inviscid, incompressible flow in the $x-z$ plane with  piecewise linear background velocity profile $\bar{U}(z)$ (parallel to the streamwise direction $x$) and piecewise constant background density profile $\rho(z)$; see figure \ref{fig:schematic_full}. Unless otherwise stated, the vertical axis $z$ points upwards, i.e.\, opposite to the direction gravity. In this system there are $N$ material interfaces $z=\eta_1,\,\eta_2,\ldots \eta_N$, which are material curves across which at least one of the flow variables - background density, vorticity (defined as $\Omega =  - d \bar{U}/d z $), or velocity, may be discontinuous. In figure \ref{fig:schematic_full} we have density discontinuity across all  interfaces. Density stratification is always assumed to be stable, i.e.\ $\rho_0 \leq \rho_1 \leq \rho_2 \leq \ldots \leq \rho_N$. The interface at $z=\eta_2$ is special; there is a discontinuity in $\bar{U}$, hence that interface is a \emph{vortex sheet} by definition. Such an interface can become unstable and yields KH instability \cite[]{drazin2004}. However, all other interfaces are also vortex sheets since discontinuity in either $\Omega$ or $\rho$ yields discontinuity in the perturbation velocity's tangential component 
\cite[]{caul1994,bain1994}.  Below we  obtain the evolution equation of vortex sheet strength for  $\Omega$ and/or $\rho$ discontinuity under a very generalized setting. 



\subsection{Dynamics of an interface in presence of shear - the ``shear-modified'' Bernoulli's equation}
For simplicity we focus on a single interface across which both $\Omega$ and $\rho$ are  discontinuous (e.g.\ $z=\eta_1$ in figure \ref{fig:schematic_full}). Since $\bar{U}$ is assumed continuous and piecewise linear, $\Omega$ is piecewise constant. Under such circumstances the perturbed flow is  \emph{irrotational} in each layer\footnote{Here we emphasize that for irrotationality of the  perturbed flow, it is not mandatory for $\Omega$   to be identically zero in each layer. $\Omega$ being piecewise constant also leads to irrotational perturbed flow in each layer.}. Hence we can  integrate the Euler's equation in each layer, just above and below the interface, to obtain
\begin{eqnarray}
\left[ \frac{\partial \phi'_0}{\partial t}+\frac{1}{2}  \left(u_{0}^{'2} + w_{0}^{'2}\right) -\Omega_0\psi'_0+g\eta_1\right]+\frac{P^{\prime}_0}{\rho_0}=0,
\label{eq:top1}\\
\left[ \frac{\partial \phi'_1}{\partial t}+\frac{1}{2}  \left(u_{1}^{'2} + w_{1}^{'2}\right) -\Omega_1\psi'_1+g\eta_1\right]+\frac{P^{\prime}_1}{\rho_1}=0.\,
\label{eq:bottom1}
\end{eqnarray}
Here $g$ is the acceleration due to gravity and $\rho_i$, $\phi'_i$, $u'_i$, $w'_i$, $\Omega_i$, $\psi'_i$ and $P'_i$ are respectively the background density, perturbation velocity potential (defined since perturbation velocity is irrotational), perturbation horizontal velocity, perturbation vertical velocity (same as total vertical velocity), background vorticity, perturbation streamfunction and perturbation pressure in the $i$-th layer, where $i=\,0$ or $1$. 
In absence of background shear, (\ref{eq:top1}) and (\ref{eq:bottom1}) basically represent unsteady Bernoulli's equation. The extra term, $-\Omega_i \psi'_i$, appears due to the presence of background shear.  To the best of our knowledge, (\ref{eq:top1})-(\ref{eq:bottom1}) is the most general description of the ``shear-modified'' Bernoulli's equation. A special case of these equations was formulated in \citet{simmen}, which considered the effect of constant shear on a free surface gravity wave (air considered as a passive fluid of zero density) in a steady frame.
The formulation of Simmen has been used in many surface gravity wave problems that includes linear shear current in the water region  and a passive air above it \cite[]{kishida1988stokes,constantin2004exact}.  

The dynamic boundary condition is usually obtained by subtracting (\ref{eq:top1}) and (\ref{eq:bottom1}) and equating the pressure, which has to be continuous across an interface in absence of surface tension. However, when surface tension is present, pressures above and below an interface are related by the Young-Laplace equation
\begin{equation}
\label{eq4}
P_{0}^\prime = P_{1}^\prime + \sigma \kappa,
\end{equation}
where $ \sigma $ is the coefficient of surface tension and $ \kappa $ is the curvature of the interface. The curvature is calculated by taking divergence of the  vector normal to the interface:
\begin{equation}
\label{eq11mis}
\kappa = -\tilde{\nabla} \cdot \mathbf{\hat{n}},
\end{equation}
where $ \tilde{\nabla} \equiv \nabla - \left(\mathbf{\hat{n}} \cdot \nabla\right) \nabla$. On subtracting (\ref{eq:bottom1}) from (\ref{eq:top1}) we get:


\begin{multline}
\left[ \frac{\partial \phi'_0}{\partial t}+\frac{1}{2}  \left(u_{0}^{'2} + w_{0}^{'2}\right) -\Omega_0\psi'_0\right]-
\left[ \frac{\partial \phi'_1}{\partial t}+\frac{1}{2}  \left(u_{1}^{'2} + w_{1}^{'2}\right) -\Omega_1\psi'_1\right]+ \frac{P_0^\prime}{\rho_0}-\frac{P_1^\prime}{\rho_1}=0.
\label{eq:dynamic_bc}
\end{multline}
Similarly, on adding (\ref{eq:top1}) and  (\ref{eq:bottom1})  we get:
\begin{multline}
\left[ \frac{\partial \phi'_0}{\partial t}+\frac{1}{2}  \left(u_{0}^{'2} + w_{0}^{'2}\right) -\Omega_0\psi'_0+g\eta_1\right]+\\
\left[ \frac{\partial \phi'_1}{\partial t}+\frac{1}{2}  \left(u_{1}^{'2} + w_{1}^{'2}\right) -\Omega_1\psi'_1+g\eta_1\right]+ \frac{P_0^\prime}{\rho_0}+\frac{P_1^\prime}{\rho_1}=0.
\label{eq:dynamic_bc2}
\end{multline}

Our objective is to find the time evolution of vortex sheet strength $\gamma$, which is defined as follows:
\begin{equation}
\gamma \equiv \mathbf{\left(u'_0-u'_1\right)}\cdot\bf{\hat{s}},
\end{equation}
where $ \mathbf{u'_0}=(u'_0,w'_0)$, $\mathbf{u'_1}=(u'_1,w'_1)$ and $\mathbf{\hat{s}}$ is the unit tangent vector to the interface. 

For converting the formulation to velocities instead of potentials, we operate $\nabla$ on (\ref{eq:dynamic_bc}), then project it along the tangential direction (by taking an inner product  with $\mathbf{\hat{s}}$) to obtain
\begin{equation}
\left\lbrace\left[ \frac{D_0\textbf{u}'_0}{D t} -\Omega_0\nabla\psi'_0\right] -
\left[ \frac{D_1\textbf{u}'_1}{D t} -\Omega_1\nabla\psi'_1\right]+ \frac{\nabla P_0^\prime}{\rho_0}-\frac{\nabla P_1^\prime}{\rho_1}\right\rbrace\cdot \mathbf{\hat{s}}=0.
\label{eq:intermediate1}
\end{equation}
Performing the same operation on (\ref{eq:dynamic_bc2}) yields 
\begin{equation}
\left\lbrace\left[ \frac{D_0\textbf{u}'_0}{D t} -\Omega_0\nabla\psi'_0 + g\mathbf{\hat{k}}\right] +
\left[ \frac{D_1\textbf{u}'_1}{D t} -\Omega_1\nabla\psi'_1 + g\mathbf{\hat{k}}\right]+ \frac{\nabla P_0^\prime}{\rho_0}+\frac{\nabla P_1^\prime}{\rho_1}\right\rbrace\cdot\mathbf{\hat{s}}=0.
\label{eq:intermediate2}
\end{equation}
Here  $ D_{i}/Dt \equiv \partial/\partial t + \mathbf{u'_i}\cdot\nabla $ is the (perturbation) material derivative in the $i^{\mathrm{th}}$ layer;\,$i=0,1$.  Furthermore $\nabla = \mathbf{\hat{s}}\,\partial/\partial s$,  where $s$ is the arc-length coordinate. Eliminating the pressure terms in  (\ref{eq:intermediate1}) and (\ref{eq:intermediate2}) and using (\ref{eq4}) we obtain the final evolution equation of vortex sheet strength: 
\begin{multline}
\label{eq:evol}
\frac{d \gamma}{d t} = 2At\frac{d \textbf{U}^\prime}{d t}\cdot \mathbf{\hat{s}} + \frac{\beta + At}{4}\left(\frac{\partial \gamma^2}{\partial s}\right) - (1 + \beta At)\gamma \frac{\partial \textbf{U}^\prime}{\partial s}\cdot \mathbf{\hat{s}} + 2At g\frac{\partial{\eta_1}}{\partial{s}} - \frac{\sigma}{\rho_{av}}\frac{\partial \kappa}{\partial s} \\ - \left(At-1\right)\Omega_{0}\nabla \psi_{0}^\prime\cdot \mathbf{\hat{s}} -\left(At+1\right)\Omega_{1}\nabla \psi_{1}^\prime\cdot \mathbf{\hat{s}}. 
\end{multline}
Here $ \rho_{av}=(\rho_0 + \rho_1)/2 $ is the average density, $At=(\rho_1-\rho_0)/(\rho_1+\rho_0)$ is the Atwood number and $\mathbf{U'}=(\mathbf{u'_0}+\mathbf{u'_1})/2$ is the average perturbation velocity at the interface. The term $2At g\partial \eta_1/\partial s$ denotes the baroclinic generation of vorticity.  The operator 
$d/dt \equiv \partial/\partial t + \mathbf{q}\cdot \nabla$ denotes (perturbation) material derivative following the interface, where $\mathbf{q}$ is the  interfacial perturbation velocity defined as follows:
\begin{equation}
\label{eq:int_vel}
\mathbf{q} \equiv \mathbf{U'} + \frac{1}{2}\gamma \beta \mathbf{\hat{s}}.
\end{equation}
In the above equation, $\beta=[-1,1]$ is a weight parameter such that $\beta=1\,(-1)$ implies that the interface is moving with the upper (lower) fluid. We have chosen $\beta=0$ in this study, implying that the interface travels with the average velocity of the two layers.
 
The vortex sheet evolution equation, (\ref{eq:evol}), is in its most general form. The last two terms of it denote the effect of piecewise linear background velocity (or piecewise continuous background vorticity). We would emphasize here that these terms were absent in the previous derivations of vortex sheet evolution; see  (2.15) of \citet{baker1982generalized}, (15) of \citet{tryggvason1988numerical}, or (9) of \citet{sohn2010long}, since these authors chose the base flow to be irrotational. Furthermore, our derivation strategy has also been different - we have obtained  (\ref{eq:evol}) from unsteady Bernoulli's equation modified by background shear; see (\ref{eq:top1})-(\ref{eq:bottom1}). 


\subsection{Evolution of multiple interfaces - the modified Birkhoff-Rott equation}


The interface variables can be parameterized using the arc-length coordinate $s$. The average perturbation velocity $\textbf{U}^\prime$ of an infinitely long interface, which is also a vortex sheet, can be evaluated  using the Biot-Savart integral:
\begin{equation}
\textbf{U}'\left(s,t\right)= -\frac{1}{2\pi} 
\,\mathrm{P.V.} \int_{-\infty}^{+\infty}\frac{\mathbf{\hat{j}}\times\left(\textbf{X}\left(s,t\right)-\textbf{X}\left( \tilde{s},t\right) \right)}{|\textbf{X}(s,t)-\textbf{X}(\tilde{s},t)|^2}\gamma(\tilde{s},t)\,d\tilde{s}.
\end{equation}
Here \textbf{X} is the position vector representing an arbitrary point on the interface, and is given by $\textbf{X} = x(s,t)\mathbf{\hat{i}}+z(s,t)\mathbf{\hat{k}}$.   P.V.\, in front of the integral implies principal value  (hereafter, we don't put P.V. in front of Biot-Savart integrals, assuming it is expected), and $ \mathbf{\hat{j}} $ is the unit vector perpendicular to the plane containing the vortex sheet.

The Biot-Savart law for periodic boundary condition (as is present in the interfaces we consider) reduces to the Birkhoff-Rott equation:
\begin{equation}
u'-i w' = \frac{i}{2\lambda}\int_{0}^{\lambda}\tilde{\gamma}\cot\bigg[\frac{\pi(\chi - \tilde{\chi})}{\lambda}\bigg] d\tilde{s}.
\end{equation}
Here $ u' $,\,$ w' $ are the horizontal and vertical components of $\mathbf{U'}$, $ \chi $ is the complex position of the interface: $ \chi = x(s,t) + i z(s,t) $, $ \lambda $ is the wavelength, and $\tilde{\gamma} \equiv \gamma (\tilde{s},t)$. 
In this paper we have extended the Birkhoff-Rott equation to multiple interfaces, which can be  mathematically written as follows:
\begin{equation}
\label{eq15}
u'_{l}-i w'_{l} = \sum_{k=1}^{M}\frac{i}{2\lambda}\int_{0}^{\lambda}\tilde{\gamma}_{k}\cot\bigg[\frac{\pi(\chi_{l} - \tilde{\chi}_{k})}{\lambda}\bigg]\,d\tilde{s}_{k}.
\end{equation}
Here we are considering a system with $M$ interfaces; the indices $ k $ and $ l $ represent the $ k^{\mathrm{th}} $ and $ l^{\mathrm{th}} $ interfaces respectively.

\subsection{Numerical technique}
\label{sec:num_tech}
An efficient and accurate numerical method for solving the Biot-Savart or the Birkhoff-Rott equation is by discretizing each interface into a series of point vortices. These vortices are Lagrangian markers, given $\gamma$ at a particular time $t$ an interface evolves following
\begin{equation}
\label{eq:int_update}
\frac{d\mathbf{X}}{dt} = \mathbf{q}+U_B\mathbf{\hat{i}},
\end{equation}
where $U_B$ is the background velocity. In general we have $U_B=\bar{U}(z)$, i.e. $U_B$ does not evolve in time. However, if contour dynamics is implemented (e.g.\  the plunging breaker problem in  \ref{sec:plung_break}), then $U_B$ would evolve according to (\ref{eq23}).

Since we are using Lagrangian formalism, \textbf{q} and $\mathbf{U'}$ are both Lagrangian velocities, which are related by (\ref{eq:int_vel}). The average perturbation velocity $\mathbf{U'}$ can be obtained from the discretized version of (\ref{eq15}), i.e.\   Birkhoff-Rott equation  for multiple interfaces:
\begin{eqnarray}
u'_{il} = \frac{1}{2\lambda}\sum_{k=1}^{M}\sum_{j=1, j \neq i}^{N}\Gamma_{jk}\frac{\sinh [\alpha(z_{il}-z_{jk})]}{\cosh [\alpha(z_{il}-z_{jk})] - \cos [\alpha(z_{il}-z_{jk})] + \delta^2}, \label{eq:birkhoff_disc_u}\\
w'_{il} = -\frac{1}{2\lambda}\sum_{k=1}^{M}\sum_{j=1, j \neq i}^{N}\Gamma_{jk}\frac{\sin[ \alpha(x_{il}-x_{jk})]}{\cosh[ \alpha(z_{il}-z_{jk})] - \cos[ \alpha(z_{il}-z_{jk})] + \delta^2}. \label{eq:birkhoff_disc_w}
\end{eqnarray}
Here $ u'_{il}$ and $w'_{il} $ respectively denote the horizontal and vertical components of the average perturbation velocity $\mathbf{U'_{il}}$ induced by the $i^{\mathrm{th}} $ point vortex located at the $ l^{\mathrm{th}} $ interface. The wavenumber $ \alpha = 2 \pi/\lambda$;  $ \Gamma_{jk} $ denotes the circulation strength of the $ j^{\mathrm{th}}$ point  vortex located at the $ k^{\mathrm{th}} $ interface. Similarly $ x_{il}$ and $x_{jk} $ respectively denote the $ x $ coordinates of the $ i^{\mathrm{th}} $ and $ j^{\mathrm{th}} $ point vortices located at the $ l^{\mathrm{th}} $ and $ k^{\mathrm{th}} $ interfaces. The same notation applies for the $ z $ coordinates. The circulation strength of point vortices is given by
\begin{equation}
\Gamma_{jk} = \gamma_{jk} \Delta s_{jk},
\end{equation}
where $ \Delta s_{jk} $ is the arc-length about the $j^{\mathrm{th}}$ point  vortex  at the $ k^{\mathrm{th}} $ interface, and is given by
\begin{equation}
\Delta s_{jk} = \frac{1}{2}\sqrt{ (x_{j+1,k} - x_{j-1,k})^2 + (z_{j+1,k} - z_{j-1,k})^2}.
\end{equation}
Point vortices being Lagrangian parcels can cluster in certain region(s) of the interface, which may lead to singularities. Such issues can be avoided by introducing a desingularization parameter $\delta$, known as the Krasny parameter \cite[]{krasny1986desingularization}, in the denominators of  (\ref{eq:birkhoff_disc_u}) and (\ref{eq:birkhoff_disc_w}). 

Once each interface moves to its new location following (\ref{eq:int_update}) in a time interval $\Delta t$, their vortex sheet strengths are evolved in time according to (\ref{eq:evol}). For solving these equations, we require initial conditions for vortex sheet strength and initial shape of the interfaces.  Once the initial shape is provided, the initial velocities of the point vortices are found using  (\ref{eq:birkhoff_disc_u})-(\ref{eq:birkhoff_disc_w}). 
Since vortex strength and velocities are both coupled together we will need to solve for them iteratively. For this we have followed the method outlined in \citet{sohn2010long}.  
We have also followed \citet{sohn2010long} for handling resolution issues of the vortex sheets. Unless otherwise stated, in all our calculations we initially discretize one wavelength ($2\pi$ distance) of a vortex  sheet by $257$ point vortices. 
As mentioned previously, with the progress of time, an interface can develop poor resolution in certain areas and clustering in others. At late times, point clustering and diverging occur irregularly
along the interfaces. To deal with this issue we use an insertion-deletion scheme. We define a threshold arc-length $ \Delta s_{\mathrm{thresh}} $, if arc-length at any location $ i $ becomes less than the threshold, i.e.\ $ \Delta s_{ik}<\Delta s_{\mathrm{thresh}} $, then we delete that vortex; vice-versa we insert a point vortex. The $x$ coordinate of the inserted vortex $ \left(x_{p},y_{p}\right) $ is taken to be the average of the $x$ coordinates of $i^{\mathrm{th}}$ and $(i+1)^{\mathrm{th}}$ vortices: $ x_{p} = \left(x_i + x_{i+1}\right)/2 $; likewise we interpolate for the $ z $ coordinate and the vortex strength $\gamma_p$. We have validated our code with the published results of \citet{baker1980vortex} and \citet{sohn2010long}. Our code is found to accurately replicate the late time KH roll-up, as well as surface gravity wave breaking.

\subsection{Modeling fixed boundary as a vortex sheet}
\label{sec:fixed_bc}
Up to this point we have discussed the mathematical formulation of interaction between free interfaces, which are basically vortex sheets. By free interface we mean interfaces that can freely evolve. Examples of such interfaces include air-sea interface, the fresh water - salt water interface (pycnocline). However, a bottom topography, e.g.\ the sea bed, is also an interface. The difference is that it is fixed (cannot evolve freely in time). The key point to note here is that a \emph{fixed interface is also a vortex sheet}, and therefore can be understood following similar mathematical principles discussed earlier. The reason behind a bottom topography acting like a vortex sheet can be understood as follows. Above the  topography (which satisfies free-slip boundary condition owing to the fact that the fluid is inviscid), the fluid has a velocity (that is generated by the point vortices on the free interfaces), while below the topography the velocity has to be zero. Hence there is a velocity discontinuity across the topography, making it a vortex sheet. The prime differences between a free and a fixed interface are that for the latter (i) the normal velocity has to be zero (impenetrability condition), and  (ii) there is no separate evolution equation for the vortex sheet strength. The vortex sheet strength of a fixed interface evolves with the vortex sheet strength of the free interfaces, hence the problem must be solved in an iterative fashion.  Our formulation in this regard is similar, but simpler than the formulation used by \citet{baker1982generalized} for simulating breaking nonlinear surface gravity waves.

\subsubsection{Mathematical formulation}
\label{subsub:fixed_int}
Our system can be composed of multiple free interfaces and multiple fixed interfaces (i.e.\ boundaries). The evolution of the vortex sheet strength of free interfaces is governed by  (\ref{eq:evol}). 
The vortex strength of the fixed bottom is evaluated using the fact that the velocity below the interface is zero and  the velocity above the interface is only present in the direction tangential to the bottom interface:
\begin{equation*}
\gamma_b = \left(\mathbf{u'_0}-\mathbf{u'_1}\right)\cdot{\hat{\textbf{s}}} =\textbf{u}_0^\prime\cdot{\hat{\textbf{s}}},
\end{equation*}
where $ \bf{u'_0} $ is the perturbation velocity above the fixed interface, while $ \bf{u'_1}=\bf{0} $ is the perturbation velocity below it.  Since the Birkhoff-Rott equation evaluates the average perturbation velocity at the interface we have 
\begin{equation*}
\mathbf{U'} = \frac{\mathbf{u'_0} + \mathbf{u'_1}}{2} = \frac{\mathbf{u'_0}}{2}.
\end{equation*}
Hence the vortex strength of the fixed interface becomes
\begin{equation}
\label{eq:evol_bottom}
\gamma_{b} = 2\mathbf{U'}\cdot{\hat{\textbf{s}}}.
\end{equation}
The principal vortex velocity $\textbf{U}^\prime$ is itself calculated from the vortex strengths of both free and fixed interfaces, and is given by (in complex representation):
\begin{equation}
\label{eq:birkhoff_bottom}
u'_{b}-i w'_{b} = \frac{i}{2\lambda}\int_{0}^{\lambda}\tilde{\gamma}_{f}\cot\bigg[\frac{\pi(\chi_{b} - \tilde{\chi}_{f})}{\lambda}\bigg]d\tilde{s}_{f} + \frac{i}{2\lambda}\int_{0}^{\lambda}\tilde{\gamma}_{b}\cot\bigg[\frac{\pi(\chi_{b} - \tilde{\chi}_{b})}{\lambda}\bigg]d\tilde{s}_{b}.
\end{equation}
Here the subscripts $ f $ and $ b $ denote the free interface and the fixed bottom respectively. Since the equation for the vortex strength of the bottom interface is implicit in nature, we  solve for it iteratively.

\subsubsection{Numerical Technique}
The numerical technique for solving the system of equations governing the interactions between free and fixed interfaces is similar to that discussed in  section \ref{sec:num_tech}. The system of governing equations (\ref{eq:evol}), (\ref{eq15}),  (\ref{eq:int_update}), (\ref{eq:evol_bottom}) and (\ref{eq:birkhoff_bottom}) represent an initial value problem and hence the knowledge of the initial shape of the interfaces and their vortex strengths is imperative. The only difference here is that at every time level we iterate until  the acceleration of the free surface and vortex sheet strength of the bottom converge. The vortex sheet strength of the bottom is updated using equation (\ref{eq:evol_bottom}).

\section{Flows with finite shear - surface wave breaking problems}
\label{sec:finite_shear}
\subsection{Spilling breaker}

\begin{figure}
\centering
\includegraphics[width=0.8\linewidth]{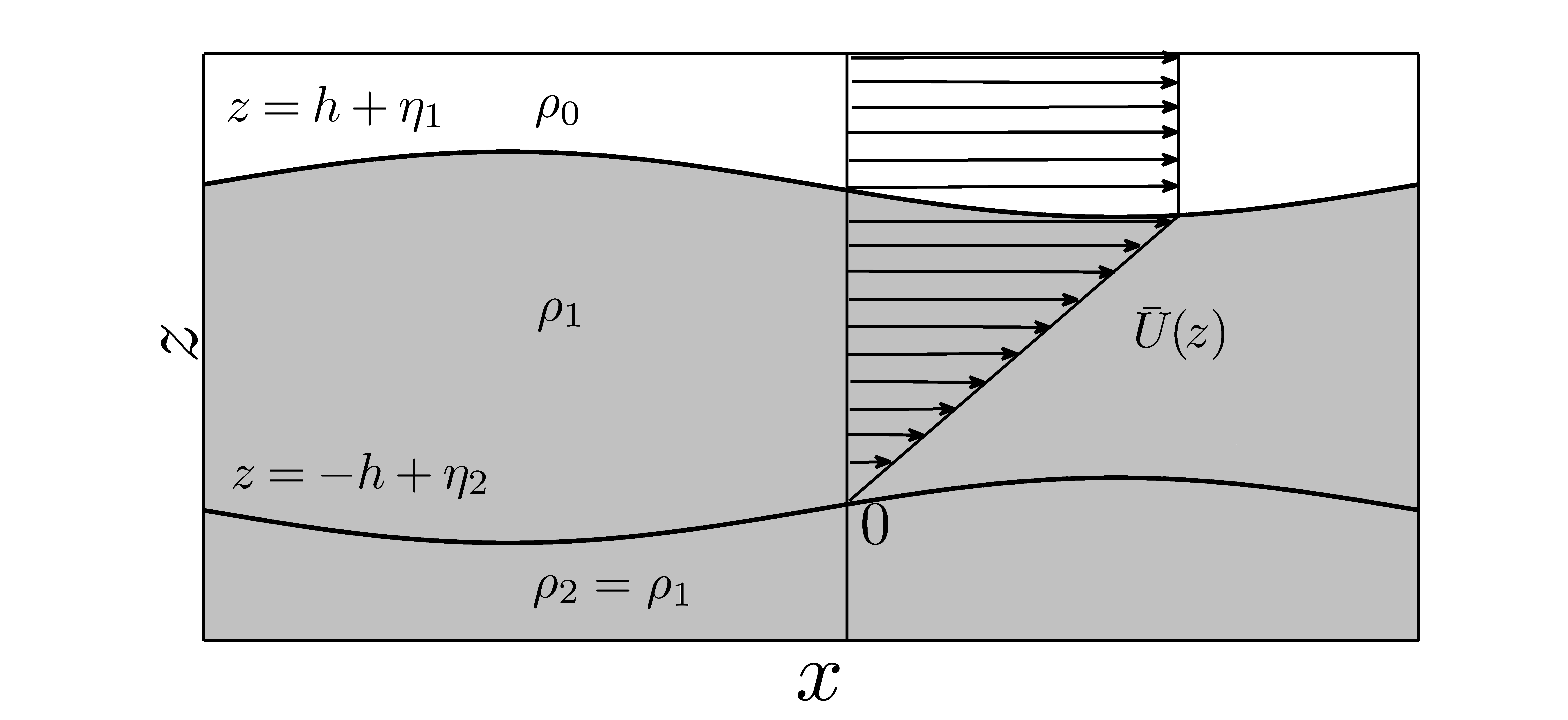}
\captionof{figure}{A schematic showing the initial configuration for a spilling breaker. The white region denotes air while the gray region denotes water.}
\label{spilling_schematic}
\end{figure}

Deep water gravity wave breaking is characterized as either spilling or plunging. In a plunging breaker the wave grows, steepens, forms a horizontal jet which later falls under the effect of gravity onto the front face of the crest of wave. In numerical simulations of a plunging breaker the wave is forced by an asymmetric pressure distribution. \citet{longuet1976deformation} did simulations of a plunging breaker using boundary integral methods. A spilling breaker is different in that the wave grows, becomes nonlinear, then a bulge  forms at the crest of the wave, and small capillary waves originate at the toe of this bulge. A theoretical model of spilling breaker was proposed by \citet{longuet1994shear}, we have used this model for our simulations. In this section we present a detailed description of the initial conditions and then present the numerical simulation results. The initial condition derivation for the system is quite general in the sense that the air-water  interface (located about $z=h$) has  density jump, vorticity jump, as well as surface tension. A schematic of the system is shown in figure \ref{spilling_schematic}. The interface (located about $z=-h$) has only a vorticity jump. The base vorticity and density profiles  are given by


\begin{equation}
\Omega(z) = \left\{
        \begin{array}{cc}
        \Omega_0 & \quad h \leq z \\ 
        \Omega_1 & \quad  -h \leq z < h \\ 
        \Omega_2 & \quad z<-h
        \end{array}
    \right.
   \qquad\qquad \bar{\rho}(z) = \left\{
        \begin{array}{cc}
        \rho_0 & \quad h \leq z \\ 
        \rho_1 & \quad  -h\leq z< h\\ 
        \rho_2 & \quad z<-h.
        \end{array}
    \right.
\end{equation}
For a spilling breaker $\Omega_0=\Omega_2=0$ and $\Omega_1$ is a constant base shear. Furthermore $\rho_0=0$ (density of air assumed to be zero) and $\rho_2=\rho_1$. However we will proceed without making these assumptions at this point so as to arrive at a general initial condition.

\subsubsection{Initial conditions}
 Following (\ref{eq:top1})-(\ref{eq:bottom1}),  linearized Bernoulli's equations (in the presence of shear) above and below the interfaces are given by
\begin{eqnarray}
P_{j}^{\prime\,(1)} = -\rho_{j}\left[\frac{\partial \phi'_{j}}{\partial t} + U_{1}\frac{\partial \phi'_{j}}{\partial x}+g\eta_{1}- \Omega_{j}\psi'_{j}\right],\,\,j = 0,\,1,\\
P_{j}^{\prime\,(2)} = -\rho_{j}\left[\frac{\partial \phi'_{j}}{\partial t} + U_{2}\frac{\partial \phi'_{j}}{\partial x}+g\eta_{2}- \Omega_{j}\psi'_{j}\right],\,\,j = 1,\,2.\,
\end{eqnarray}
Note here that the subscript $ j = 0, 1, 2 $ refers to the different fluid layers in our three-layered system. Furthermore, $P_{j}^{\prime\,(k)}$ implies perturbation pressure  of the $j^{\mathrm{th}}$  layer in the neighborhood of the $k^{\mathrm{th}}$ interface. Also $U_1 \equiv \bar{U}(z=h)$ and $U_2 \equiv \bar{U}(z=-h)$ are basically the background velocities at the respective interfaces.  Applying normal stress balance at the interfaces, we obtain
\begin{eqnarray}
P_{0}^{\prime\,(1)}=P_{1}^{\prime\,(1)}+\sigma \kappa,\\
P_{1}^{\prime\,(2)}=P_{2}^{\prime\,(2)},
\end{eqnarray}
where $\sigma$ is the surface tension coefficient at the air-water interface. The perturbation velocity potentials satisfy the Laplace equation
\begin{equation}
\nabla^2 \phi'_{j} = 0.
\end{equation}
Now we assume normal mode forms for all the perturbation quantities:
\begin{align*}
&\phi'_{j}\left(x,z,t\right)=\tilde{\phi}_{j} \left(z\right)e^{i\left(\alpha x-\omega t\right)},\\
&\eta_{j}\left(x,t\right)=\tilde{\eta}_{j} e^{i\left(\alpha x-\omega t\right)}.
\end{align*}
On substituting the normal mode solutions into the Laplace equation we get:
\begin{align*}
\frac{d^2 \tilde{\phi}_{j}}{dz^2}-\alpha^2\tilde{\phi}_{j}=0.
\end{align*}
This implies that the eigenfunctions of the potentials in the $z$-direction are of exponential form. Using the fact that far away from the interfaces these potentials must decay to zero (evanescent condition), we can eliminate a few terms. Finally we are left with four constants appearing in eigenfunctions of velocity potentials and the two interfacial displacements ($ \tilde{\eta_{1}}  $ and $ \tilde{\eta_{2}}  $):
\begin{align*}
& \tilde{\phi}_{0}(z) = C_{1}e^{-\alpha z}, \\
& \tilde{\phi}_{1}(z) = C_{2}e^{-\alpha z} + C_{3}e^{\alpha z}, \\
& \tilde{\phi}_{2}(z) = C_{4}e^{\alpha z}.
\end{align*}
We need six equations to solve for this system. Hence we write four kinematic equations, just above and below each of the two interfaces, and two normal stress balance conditions across each of the interfaces.
\begin{align}
& \frac{\partial \eta_{1}}{\partial t} + U_1\frac{\partial \eta_{1}}{\partial x} = \frac{\partial \phi'_{k}}{\partial z};\,\,  k=0,\,1, \\
& \frac{\partial \eta_{2}}{\partial t} + U_2\frac{\partial \eta_{2}}{\partial x} = \frac{\partial \phi'_{k}}{\partial z};\,\, k=1,\,2. 
\end{align}
 Finally we formulate the problem as a system of linear homogeneous equations $ \mathcal{A}\mathcal{X} = 0 $.
The  matrix $ \mathcal{A} $ in the most general form is given by
\[
   \mathcal{A}=
  \left[ {\begin{array}{cccccc}
   -\alpha e^{- \alpha h} & 0 & 0 & 0 & i( \omega -  U_{1}\alpha) & 0 \\
   0 & -\alpha e^{-\alpha h} & \alpha e^{\alpha h} & 0 & i( \omega -  U_{1}\alpha) & 0 \\
    0 & -\alpha e^{\alpha h} & \alpha e^{-\alpha h} & 0 & 0 & i( \omega -  U_{2}\alpha) \\
    0 & 0 & 0 & \alpha e^{-\alpha h} & 0 & i( \omega -  U_{2}\alpha) \\
    A_{51} & A_{52} & A_{53} & 0 & (\rho_{0}-\rho_{1})g-\sigma\alpha^2 & 0 \\
   0 & A_{62} & A_{63} & A_{64} & 0 & (\rho_{1}-\rho_{2})g 
  \end{array} } \right],
\]
where 
\begin{align*}
& & A_{51} = i\rho_{0} e^{-\alpha h}(-\omega + U_{1}\alpha + \Omega_{0})
,\, A_{52} = -i\rho_{1} e^{-\alpha h}(-\omega + U_{1}\alpha + \Omega_{1})
,\\
& & A_{53} = -i\rho_{1} e^{\alpha h}(-\omega + U_{1}\alpha - \Omega_{1})
 ,\,A_{62} = i\rho_{1} e^{\alpha h}(-\omega + U_{2}\alpha + \Omega_{1})
 ,\\
 & & A_{63} = i \rho_{1} e^{-\alpha h}(-\omega + U_{2}\alpha - \Omega_{1})
 ,\,A_{64} = -i\rho_{2} e^{-\alpha h}(-\omega + U_{2}\alpha - \Omega_{2}).
\end{align*}
The variable vector $ \mathcal{X} $ is given by
\[
\mathcal{X} = 
\begin{bmatrix}
C_{1} & C_{2} & C_{3} & C_{4} & \tilde{\eta}_{1} & \tilde{\eta}_{2}
\end{bmatrix}^{\dagger}.
\]
On applying the condition for a non-trivial solution  we get the dispersion relation  for the system,  $ \mathcal{X} $ being  the null vector of the matrix $ \mathcal{A} $.

In case of modal instabilities  $ \omega $ obtained from the dispersion relation will be complex, and the null vector corresponding to that frequency will give us the phase locked configuration that results in exponentially growing modes. Phase locking is a state in which the relative phase between the waves does not change. This occurs when the intrinsic phase speed of the waves is counter to the base flow at the respective interfaces as well as counter to each other (counter-propagating waves)\cite{heif2005,carp2012,guha2014}. 
Now that we have solved for all the variables, the vortex sheet strength of the interfaces are given by
\begin{align}
& \gamma_{1} = \frac{\partial}{\partial x}(\phi_{0}^\prime(x,z,0) - \phi_{1}^\prime(x,z,0))\bigg\vert_{z = h}, \label{eq:gamma1}\\
& \gamma_{2} = \frac{\partial}{\partial x}(\phi_{1}^\prime(x,z,0) - \phi_{2}^\prime(x,z,0))\bigg\vert_{z = -h}. \label{eq:gamma2}
\end{align}

\subsubsection{Results}

\begin{figure}
\centering
\includegraphics[width=0.9\linewidth]{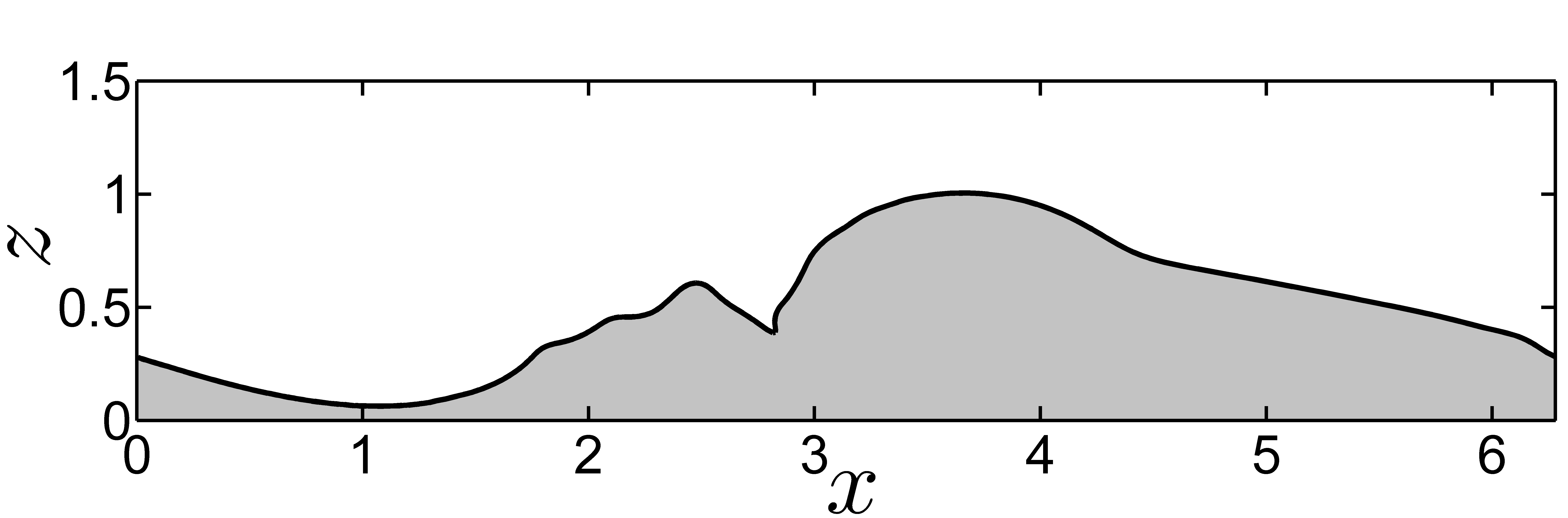}
\captionof{figure}{Nonlinear structure of a spilling breaker obtained using vortex method.}
\label{spilling_structure}
\end{figure}

The formation of spilling breaker can be understood in terms of the resonant interaction of counter-propagating waves - it forms as a result of resonance between the leftward propagating surface capillary-gravity wave (at the free surface) and the rightward propagating vorticity wave (at the vorticity interface). 
The nonlinear structure of a spilling breaker from our simulation is given in figure \ref{spilling_structure}. The values of various parameters are as follows:  $ \sigma = 0.072 $ $  \mathrm{Nm}^{-1} $, $ g = 10 $ $  \mathrm{ms}^{-2} $, $ \Omega_0 = \Omega_2 = 0 $ $  \mathrm{s}^{-1} $,   $ \Omega_1 = 10 $ $  \mathrm{s}^{-1} $, 
 $ \rho_0 = 0 $ $ \mathrm{kgm}^{-3} $, $ \rho_1 = \rho_2 = 1000 $ $ \mathrm{kgm}^{-3} $,  $ \alpha=1 $ $\mathrm{m}^{-1}$. The initial amplitude of the surface wave is $ \tilde{\eta}_1 = 10^{-3} $ m,  and $ h = 0.5 $  m.


\begin{figure}
\centering
\includegraphics[width=0.9\linewidth]{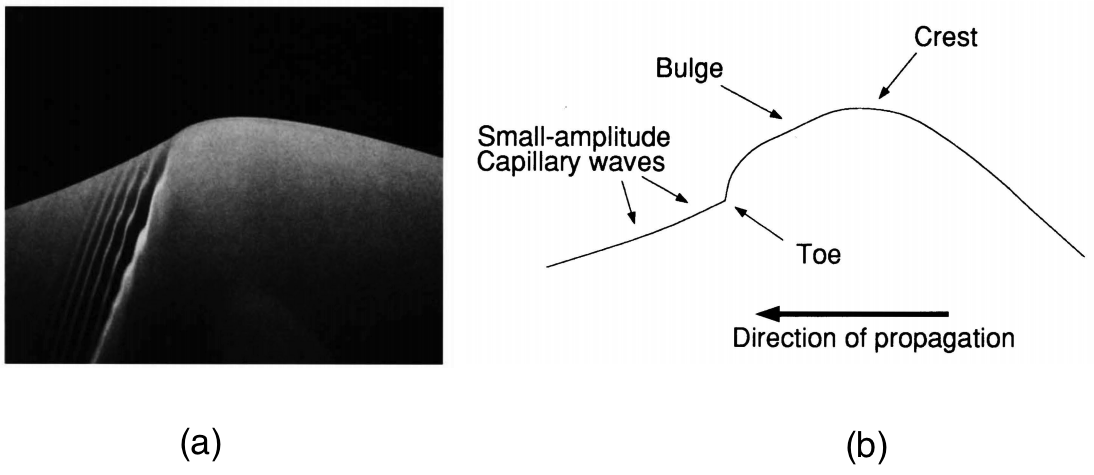}
\captionof{figure}{Nonlinear structure of a spilling breaker given in \citet{duncan1994formation}, reproduced with permission from Phys. Fluids 6, S2 (1994). Copyright 1994 American Institute of Physics. (a) Capillary wave generation and (b) schematic of the process.}
\label{spilling_structure2}
\end{figure}

The initially linear surface and interfacial waves lock in phase, resulting in an exponential growth that eventually gives rise to finite amplitude waves. These finite amplitude waves become steep at some location after which, instead of forming a horizontal jet that falls under the influence of gravity (as in the case of plunging breakers), the crest forms a bulge. At this point of time  smaller scale secondary instabilities arise near the toe of the bulge, which occur due to the vorticity shed by the primary instability, as postulated by \citet{longuet1994shear}. We can see in figure \ref{spilling_structure} that at the toe of the bulge there are some smaller scale curls which represent these secondary instabilities. Furthermore, our  simulation results show good qualitative agreement with  \citet{duncan1994formation}; see figure \ref{spilling_structure2}.

\subsection{Plunging breaker}
\label{sec:plung_break}
\begin{figure}
\centering
\includegraphics[width=1\linewidth]{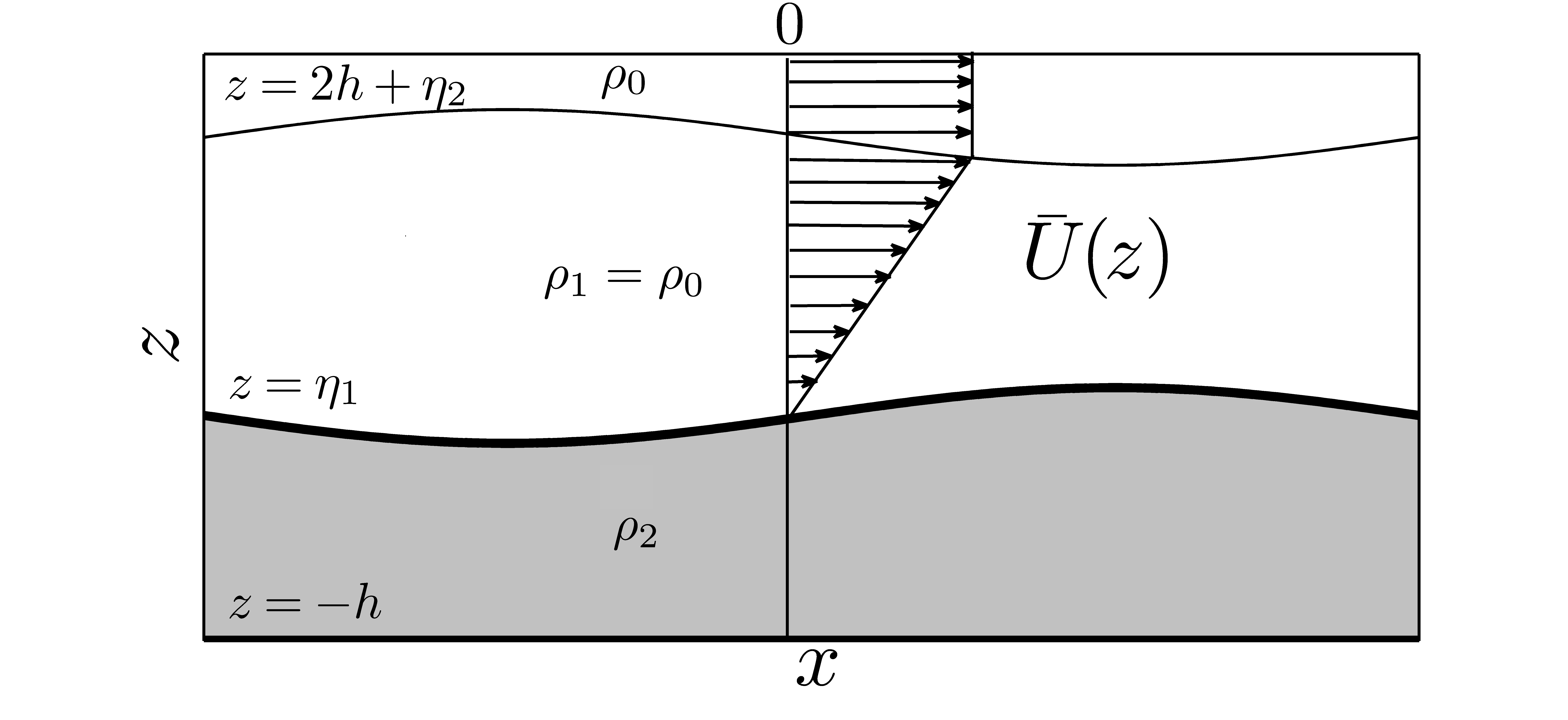}
\captionof{figure}{A schematic showing the initial condition of a plunging breaker. Here wind shear is imposed on a surface gravity wave. The white region denotes air while the gray  region denotes water. }
\label{wind_shear_schematic}
\end{figure}

In the previous section we  implemented vortex method to simulate spilling breaker, an essentially deep water phenomenon occurring at capillary scales. Here we aim to show the applicability of vortex method to capture wind shear induced breaking of long surface gravity waves - the plunging breaker phenomenon. A linear surface gravity wave in shallow water travels with a constant phase speed of $ \sqrt[]{gh} $, where $h$ is the water depth. Due to the effect of wind blowing over the ocean  surface, the wave grows exponentially until it reaches an amplitude of $\mathcal{O}(h)$,  at which point it interacts with the ocean bed and breaks. The effect of wind has been taken into account by considering the ``boundary layer'' over the free surface as a vortex patch  of constant vorticity. The schematic of the system is shown in figure \ref{wind_shear_schematic}. 


In our simulation, at every time we first evolve the vortex patch  and then we impose the velocity field due to the vortex patch on the surface gravity wave. The vortex patch evolves following contour dynamics technique\cite{pozrikidis1985nonlinear}. The velocity induced by a periodic patch of constant vorticity $\Omega$ is given by
\begin{equation}
\label{eq23}
\frac{d\textbf{X}_i}{dt} = -\frac{\Omega}{4\pi}\int_C \ln\left\lbrace\cosh\left[\alpha\left(z_i - z^\prime\right)\right] - \cos\left[\alpha\left(x_i - x^\prime\right)\right]\right\rbrace d\mathbf{x^\prime},
\end{equation}
where $ C $ denotes the contour of a single patch that is periodically repeated in the $x$ direction with wavenumber  $ \alpha$.
This interaction makes the surface wave grow exponentially. After the wave's  amplitude becomes comparable with the ocean depth, we remove the effect of the shear layer and let the wave interact with the bottom. The interaction between a surface wave and  bottom boundary has been outlined  in section \ref{sec:fixed_bc}.

The base vorticity and density profiles for this case are given by
\begin{equation}
\Omega(z) = \left\{
        \begin{array}{cc}
        \Omega_0 & \quad 2h \leq z \\ 
        \Omega_1 & \quad  0 \leq z < 2h \\ 
        \Omega_2 & \quad z<0
        \end{array}
    \right.
   \qquad\qquad \bar{\rho}(z) = \left\{
        \begin{array}{cc}
        \rho_0 & \quad 2h \leq z \\ 
        \rho_1 & \quad  0\leq z< 2h\\ 
        \rho_2 & \quad z<0.
        \end{array}
    \right.
\end{equation}
           
           
For this particular case $\Omega_0=\Omega_2=0$ and $\Omega_1$ is a constant base shear. Furthermore $\rho_0=0$ (density of air assumed to be zero) and $\rho_0=\rho_1$.

\subsubsection{Initial conditions}
\label{subsub:plung_init}
In this problem we initially start with a linear shallow water surface gravity wave which interacts with the constant vortex patch in the wind. The perturbation velocity potentials satisfy  Laplace's equation. On substituting the normal mode form, the eigenfunctions in the $z$ direction become
\begin{align*}
& \tilde{\phi_{0}}(z) = C_{1}e^{-\alpha z}, \\
& \tilde{\phi_{1}}(z) = C_{2}e^{-\alpha z} + C_{3}e^{\alpha z}. 
\end{align*}
Impenetrability at the bottom boundary yields
\begin{align*}
\frac{d \tilde{\phi_1}}{dz}\left(z\right) \bigg \vert_{z=-h}= 0.
\end{align*}
 We have four variables to solve - the three arbitrary  constants from the velocity potentials ($C_1, C_2$ and $C_3$), and the interfacial elevation $ \tilde{\eta}_1 $.  There are  two kinematic conditions, one just above and the other just below the free surface. Also we have  Bernoulli's equation across the free surface, and impenetrability condition at the bottom. Note here that we have assumed the density of air to be negligible in comparison to fluid density. So the matrix $ \mathcal{A} $ for this case becomes:
\[
   \mathcal{A}=
  \left[ {\begin{array}{cccc}
   \alpha & 0 & 0 & -i \omega \\
   0 & \alpha & -\alpha & -i \omega \\
   0 & i \rho_2 \omega & i \rho_2 \omega & -\rho_2 g \\
   0 & -\alpha e^{\alpha h} & \alpha e^{-\alpha h} & 0
  \end{array} } \right].
\]
The variable vector $ \mathcal{X} $ is as follows:
\[
   \mathcal{X}=
  \left[ {\begin{array}{cccc}
   C_1 & C_2 & C_3 &  \tilde{\eta}_1
  \end{array} } \right]^{\dagger}.
\]
We apply the condition for a non-trivial solution and obtain the dispersion relation and then find the null vector of the system. The vortex sheet strengths are given as
\begin{align}
& \gamma_{1} = \frac{\partial}{\partial x}(\phi_{0}^\prime(x,z,0) - \phi_{1}^\prime(x,z,0))\bigg\vert_{z = 0},
\label{eq:vortex_plunging1}
\\
& \gamma_{2} = \frac{\partial}{\partial x}(\phi_{1}^\prime(x,z,0))\bigg\vert_{z = -h}.
\label{eq:vortex_plunging2}
\end{align}


\begin{figure}
\centering
\includegraphics[width=0.8\linewidth]{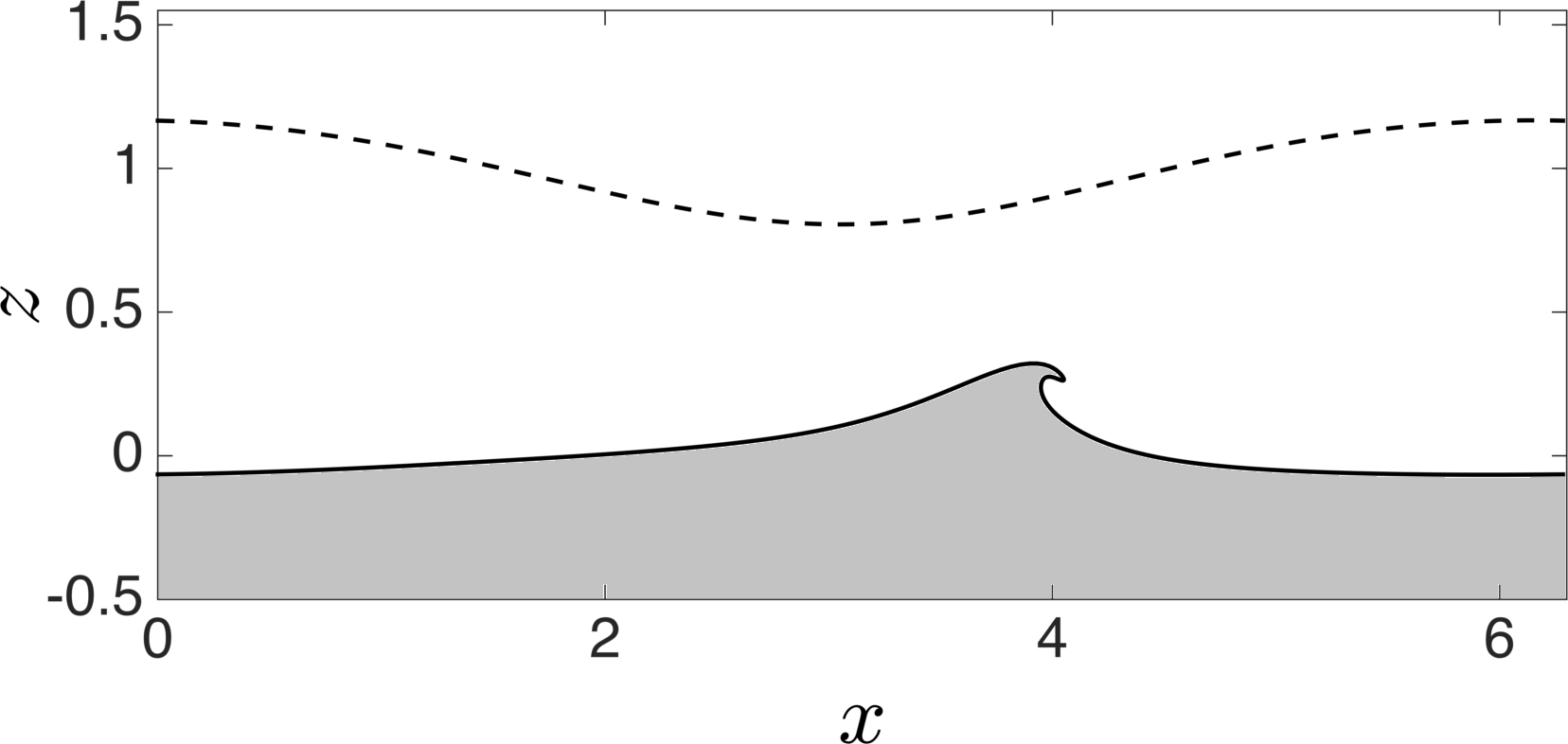}
\captionof{figure}{Plunging breaker due to imposed wind shear simulated using vortex method. The thick line represents the breaking gravity wave, while the dashed line represents the upper vorticity contour.}
\label{wind_shear_structure}
\end{figure}

\subsubsection{Results}
We start with a small amplitude surface gravity wave of $ \tilde{\eta}_1 = 10^{-3} $ m. Above it there is a vortex patch of thickness $2h$ where  $ h = 0.5 $ m, whose lower contour matches with the surface wave. The nonlinear structure that the wave forms at some later time instant is known as a plunging breaker, where initially a horizontal fluid jet is formed at the crest of the wave which then falls down under the influence of gravity. Figure \ref{wind_shear_structure} shows the structure of a plunging breaker. The values of the various parameters involved are:  $ \rho_0=0$ $\mathrm{kgm^{-3}}$, $\rho_2=1$ $\mathrm{kgm^{-3}}$, $ g = 1 $ $ \mathrm{ms}^{-2} $, $ \Omega_0 = 0 $ $ \mathrm{s}^{-1} $, $ \Omega_1 = 10 $ $ \mathrm{s}^{-1} $, $ \Omega_2 = 0 $ $ \mathrm{s}^{-1} $,  $ \alpha=1 $ $\mathrm{m}^{-1}$,  and the depth of the sea is $ d = 0.5 $ m.  These parameter values, except that of $\Omega_1$, have been taken from \citet{baker1982generalized}.  Figure \ref{wind_shear_structure} qualitatively and quantitatively matches very well with that of \citet{baker1982generalized}, even though their mechanism of nonlinear wave generation is very different from ours.

\section{Flows with finite shear - internal wave breaking problems}
\label{sec:finite_shear_internal}

\subsection{Holmboe instability}
 Holmboe instability  occurs as a result of the interaction between two counter-propagating vorticity and interfacial gravity waves \citep{holm1962,carp2012,guha2014}. In this configuration the waves phase lock with each other and show exponential growth, finally forming nonlinear structures. Holmboe instability is not as strong as KH, where complete overturning of the density interface occurs. Instead, cusp like structures  that eject from the density interface are the signature of Holmboe instability \citep{carpenter2010holmboe}.

\begin{figure}
\centering
\includegraphics[width=1.0\linewidth]{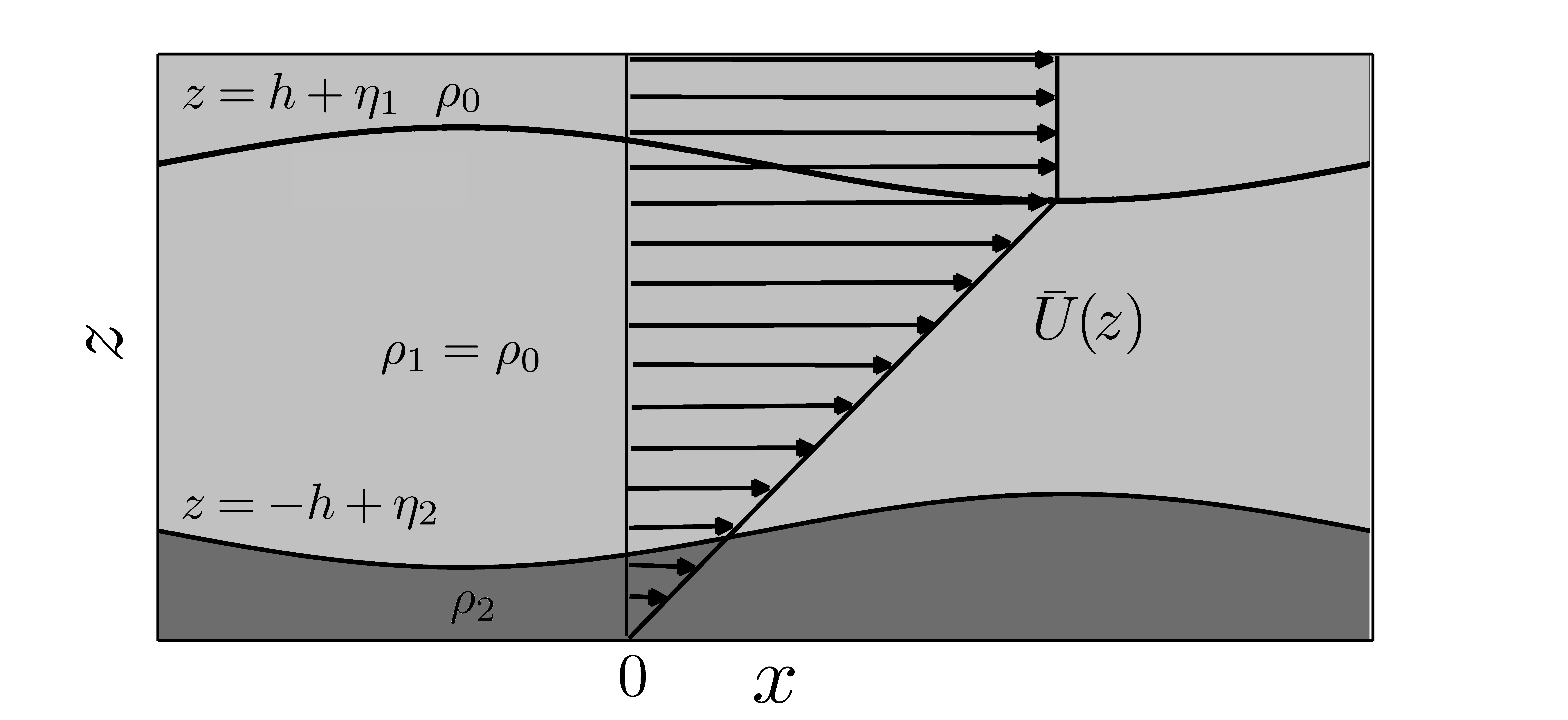}
\captionof{figure}{Schematic showing the initial configuration of Holmboe instability. The lighter gray denotes lighter while darker gray denotes heavier fluid.}
\label{holmboe_schematic}
\end{figure}

A schematic of the system has been shown in figure \ref{holmboe_schematic}. The top interface shows a vorticity discontinuity and therefore supports a vorticity wave, while the bottom interface has a density discontinuity and supports an interfacial gravity wave. The initial condition for the problem has been derived from the linear theory. The base vorticity and density profiles read
\begin{equation}
\Omega(z) = \left\{
        \begin{array}{cc}
        \Omega_0 & \quad h \leq z \\ 
        \Omega_1 & \quad  -h \leq z < h \\ 
        \Omega_1 & \quad z<-h
        \end{array}
    \right.
   \qquad\qquad \bar{\rho}(z) = \left\{
        \begin{array}{cc}
        \rho_0 & \quad h \leq z \\ 
        \rho_0 & \quad  -h\leq z< h\\ 
        \rho_2 & \quad z<-h.
        \end{array}
    \right.
\end{equation}
           
where $\Omega_0=0$ and $\Omega_1$ is a constant shear. 
           
\subsubsection{Initial conditions}
The derivation is similar to that of the spilling breaker except that here we neglect surface tension effects. So, the matrix $ \mathcal{A} $ for this case  becomes
\[
   \mathcal{A}=
  \left[ {\begin{array}{cccccc}
   -\alpha e^{-\alpha h} & 0 & 0 & 0 & i( \omega -  U_{1}\alpha) & 0 \\
   0 & -\alpha e^{-\alpha h} & \alpha e^{\alpha h} & 0 & i( \omega -   U_{1}\alpha) & 0 \\
    0 & -\alpha e^{\alpha h} & \alpha e^{-\alpha h} & 0 & 0 & i( \omega -  U_{2}\alpha) \\
    0 & 0 & 0 & \alpha e^{-\alpha h} & 0 & i( \omega -  U_{2}\alpha) \\
    A_{51} & A_{52} & A_{53} & 0 & 0 & 0 \\
   0 & A_{62} & A_{63} & A_{64} & 0 & \left(\rho_{0}-\rho_{2}\right)g
  \end{array} } \right],
\]
where we define:
\begin{align*}
& & A_{51} = i\rho_{0} e^{-\alpha h}(-\omega + U_{1}\alpha)
, A_{52} = -i\rho_{0} e^{-\alpha h}(-\omega + U_{1}\alpha + \Omega_1)
,\\ 
& & A_{53} = -i\rho_{0} e^{\alpha h}(-\omega + U_{1}\alpha - \Omega_1),
 A_{62} = i\rho_{0} e^{\alpha h}(-\omega + U_{2}\alpha + \Omega_1)
 ,\\ 
 & & A_{63} = i\rho_{0} e^{-\alpha h}(-\omega + U_{2}\alpha - \Omega_1)
 ,A_{64} = -i\rho_{2} e^{-\alpha h}(-\omega + U_{2}\alpha - \Omega_1).
\end{align*}
$U_1$ is the velocity at the upper interface while $U_2$ is the same at the lower interface. The variable vector $ \mathcal{X} $ is given by
\[
\mathcal{X} = 
\begin{bmatrix}
C_{1} & C_{2} & C_{3} & C_{4} & \tilde{\eta}_{1} & \tilde{\eta}_{2}
\end{bmatrix}^{\dagger}.
\]
On applying the condition for a non-trivial solution  we obtain the dispersion relation as well as the null vector $ \mathcal{X} $  of the matrix $ \mathcal{A} $. For exponentially growing modes, the frequencies $ \omega $ will be complex.  The  vector $ \mathcal{X} $ corresponding to that particular frequency will give us the phase locked configuration. The vortex sheet strength of the interfaces would be given by (\ref{eq:gamma1})-(\ref{eq:gamma2}).


\subsubsection{Results}
\begin{figure}
\centering
\includegraphics[width=0.75\linewidth]{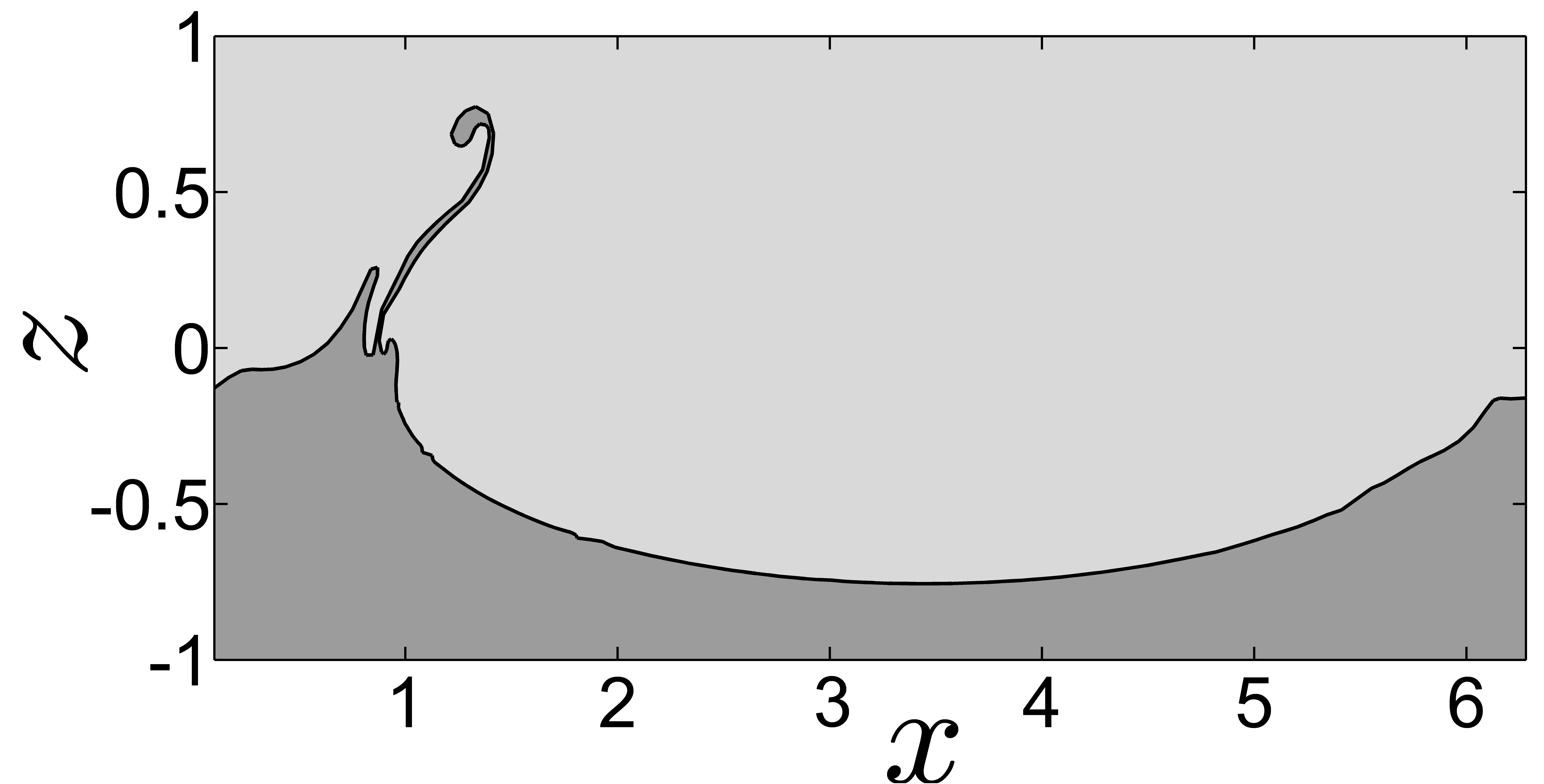}
\captionof{figure}{Vortex method used to capture the nonlinear structure of Holmboe instability. Here we show ejection associated with interfacial wave breaking.}
\label{holmboe_structure}
\end{figure}

\begin{figure}
\centering
\includegraphics[width=0.65\linewidth]{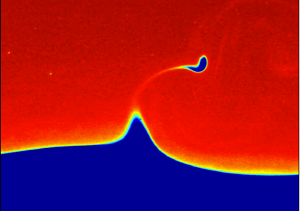}
\caption{Holmboe wave ejections obtained in  laboratory experiments (courtesy: Environmental Fluid Mechanics Lab, The University of British Columbia). Red denotes lighter (fresh water) while blue denotes heavier (salt water) fluid.}
\label{holmboe_exp}
\end{figure}

We initialize the problem with an interfacial gravity wave of  amplitude $ \tilde{\eta}_1 = 10^{-3} $ m.  The initial amplitude of the vorticity wave can be calculated from the null vector  $ \mathcal{X} $. The values of the various parameters involved are:  $ \rho_0= 0.99 $ $\mathrm{kgm^{-3}}$, $\rho_2=1$ $\mathrm{kgm^{-3}}$, $ g = 1 $ $ \mathrm{ms}^{-2} $, $ \Omega_1 = 0.2 $ $ \mathrm{s}^{-1} $, $ \alpha=1 $ $\mathrm{m}^{-1}$, and $ h = 0.5 $ m. In figure \ref{holmboe_structure} we have shown the nonlinear structure of the density interface (the lower interface in the schematic). We can see that at later times there is ejection of fluid from the interface, just as observed by \cite{carpenter2010holmboe} in their Direct numerical simulations  as well as experiments. For comparison we have shown experimental results on wave ejections associated with Holmboe instability in figure \ref{holmboe_exp}.

\subsection{Taylor-Caulfield instability}
\begin{figure}
\centering
\includegraphics[width=0.9\linewidth]{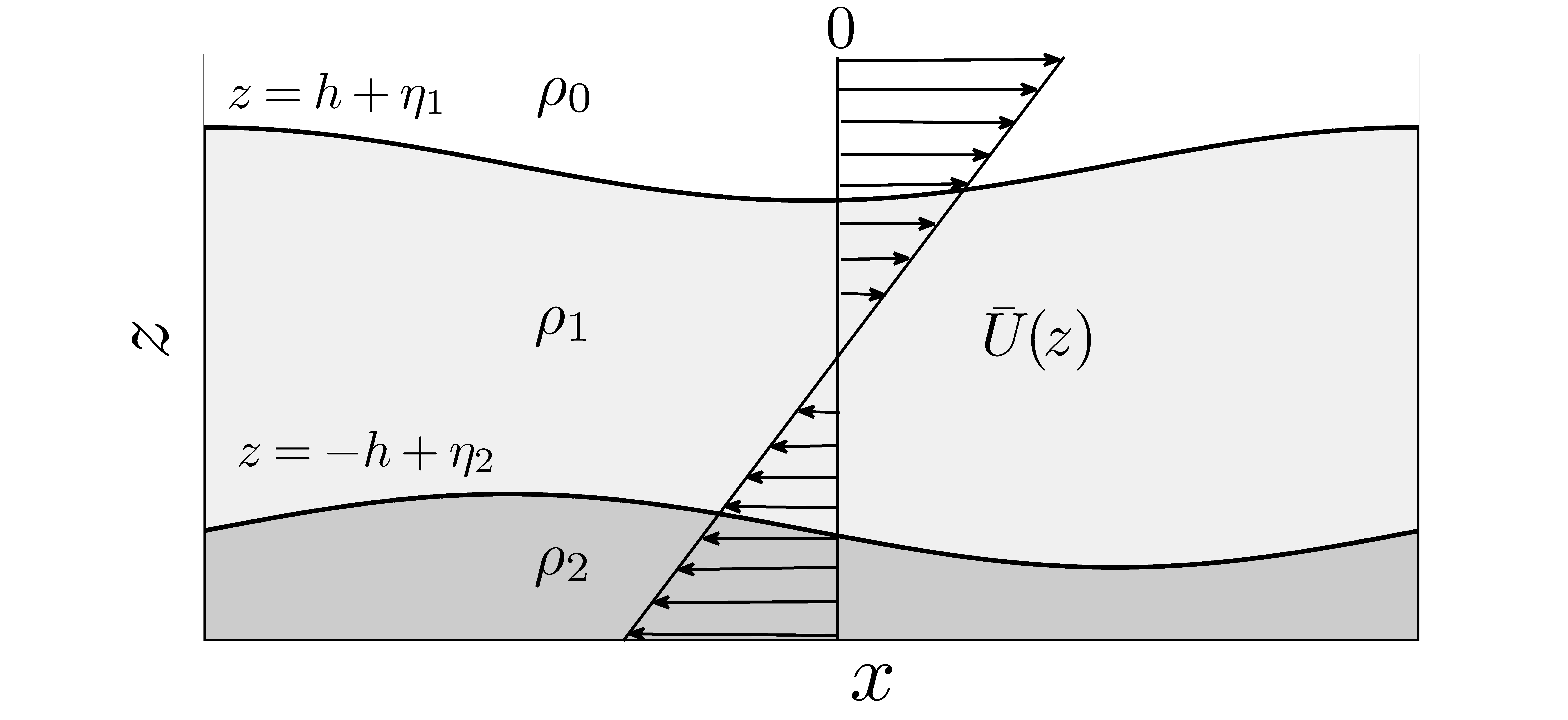}
\captionof{figure}{Schematic showing the initial configuration of Taylor-Caulfield instability. Lighter shade implies lighter and darker shade implies heavier fluid.}
\label{taylor_schematic}
\end{figure}

A stable density stratification in the presence of a linear velocity profile gives rise to unstable modes even though both the density stratification and the velocity profile are stable by themselves. \citet{taylor1931effect} proposed that an instability can arise due to the resonant interaction between two counter-propagating interfacial gravity waves, respectively located at $z=h$ and $z=-h$ in figure \ref{taylor_schematic}. These waves lock in phase and eventually give rise to nonlinear structures.

 The base vorticity and density  profiles are as follows: 
 \begin{equation}
\Omega(z) = \left\{
        \begin{array}{cc}
        \Omega_0 & \quad h \leq z \\ 
        \Omega_0 & \quad  -h \leq z < h \\ 
        \Omega_0 & \quad z<-h
        \end{array}
    \right.
   \qquad\qquad \bar{\rho}(z) = \left\{
        \begin{array}{cc}
        \rho_0 & \quad h \leq z \\ 
        \rho_1 & \quad  -h\leq z< h\\ 
        \rho_2 & \quad z<-h.
        \end{array}
    \right.
\end{equation}

           
 Here $ \Omega_0 $ is some constant base shear. 
           

\subsubsection{Initial conditions}
The derivation of initial conditions is similar to the case of Holmboe instability but here the background shear is uniform throughout. So, the matrix $ \mathcal{A} $ for this case  becomes
\[
   \mathcal{A}=
  \left[ {\begin{array}{cccccc}
   -\alpha e^{-\alpha h} & 0 & 0 & 0 & i( \omega -  U_{1}\alpha) & 0 \\
   0 & -\alpha e^{-\alpha h} & \alpha e^{\alpha h} & 0 & i( \omega -  U_{1}\alpha) & 0 \\
    0 & -\alpha e^{\alpha h} & \alpha e^{-\alpha h} & 0 & 0 & i( \omega - U_{2}\alpha) \\
    0 & 0 & 0 & \alpha e^{-\alpha h} & 0 & i( \omega -  U_{2}\alpha) \\
    A_{51} & A_{52} & A_{53} & 0 & \left(\rho_{0}-\rho_{1}\right)g & 0 \\
   0 & A_{62} & A_{63} & A_{64} & 0 & \left(\rho_{1}-\rho_{2}\right)g
  \end{array} } \right],
\]
where we define:
\begin{align*}
& & A_{51} = i\rho_{0} e^{-\alpha h}(-\omega + U_{1}\alpha + \Omega_0)
, A_{52} = -i\rho_{1} e^{-\alpha h}(-\omega + U_{1}\alpha + \Omega_0)
,\\
& & A_{53} = -i\rho_{1} e^{\alpha h}(-\omega + U_{1}\alpha - \Omega_0)
 , A_{62} = i\rho_{1} e^{\alpha h}(-\omega + U_{2}\alpha + \Omega_0)
 ,\\
 & & A_{63} = i\rho_{1} e^{-\alpha h}(-\omega + U_{2}\alpha - \Omega_0)
 ,A_{64} = -i\rho_{2} e^{-\alpha h}(-\omega + U_{2}\alpha - \Omega_0).
\end{align*}
The variable vector $ \mathcal{X} $ is given by
\[
\mathcal{X} = 
\begin{bmatrix}
C_{1} & C_{2} & C_{3} & C_{4} & \tilde{\eta}_{1} & \tilde{\eta}_{2}
\end{bmatrix}^{\dagger}.
\]
We look for complex frequencies $ \omega $ and the corresponding null vector $\mathcal{X}$ for the nonlinear structure. 
Again, the vortex strength of the interfaces are given by  (\ref{eq:gamma1})-(\ref{eq:gamma2}).
\begin{figure}
\centering
\includegraphics[width=0.8\linewidth]{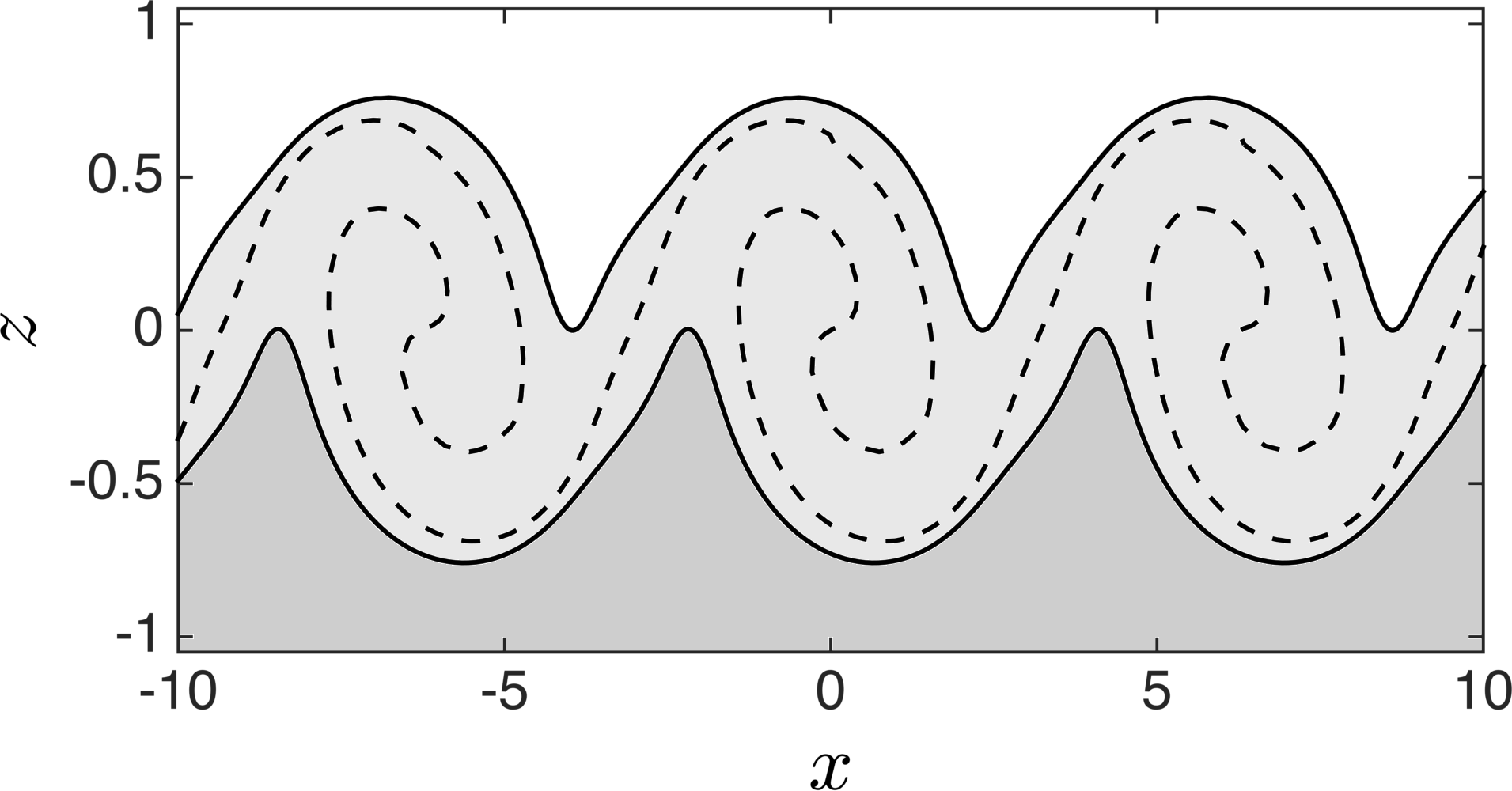}
\captionof{figure}{Nonlinear structure of Taylor-Caulfield instability obtained using vortex method. The dashed contour represents a material line.}
\label{taylor_structure}
\end{figure}

\begin{figure}
\centering
\includegraphics[width=0.6\linewidth]{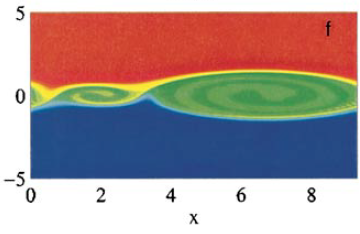}
\captionof{figure}{Nonlinear structure of Taylor-Caulfield instability obtained using DNS by \citet{lee2001nonlinear}. Red denotes lighter  while blue denotes heavier fluid. Reproduced with permission from Dynam. Atmos. Ocean 34, 103 (2001). Copyright 2001 Elsevier B.V. }
\label{taylor_exp} 
\end{figure}

\subsubsection{Results}
The values of the various parameters involved are: $ \rho_0/\rho_1 = 0.99 $, $ \rho_1/\rho_2 = 0.99 $, $ g = 1 $ $ \mathrm{ms}^{-2} $,  $\Omega_0 = 0.13 $ $ \mathrm{s}^{-1} $,  $ \alpha=1 $ $\mathrm{m}^{-1}$. The  initial amplitudes of the two waves are $ \tilde{\eta}_1 = 10^{-3} $ m and $ \tilde{\eta}_2 = 10^{-3} $ m, and  $ h = 0.5 $ m. In figure \ref{taylor_structure} we have shown the late time nonlinear structure of the Taylor-Caulfield instability. For better representation we have chosen to depict three wavelengths. The thick lines represent the envelope of the instability (looks similar to Kelvin's cat eyes); these are the late time structures of the vortex sheets shown in the schematic figure \ref{taylor_schematic}. The dashed inner contour represents a material line in the intermediate density region which displays a Kelvin-Helmholtz like roll up. Our results are qualitatively similar to \citet{lee2001nonlinear};  see figure \ref{taylor_exp}. It is to be noted that our simulation is limited to temporally unstable waves, while DNS has no such restrictions. Therefore figure \ref{taylor_exp} shows spatially growing Taylor billows.

\section{Strongly sheared flows}
\label{sec:infinite_shear}

\subsection{Jet stream instabilities}

\begin{figure}
\centering
\includegraphics[width=0.8\linewidth]{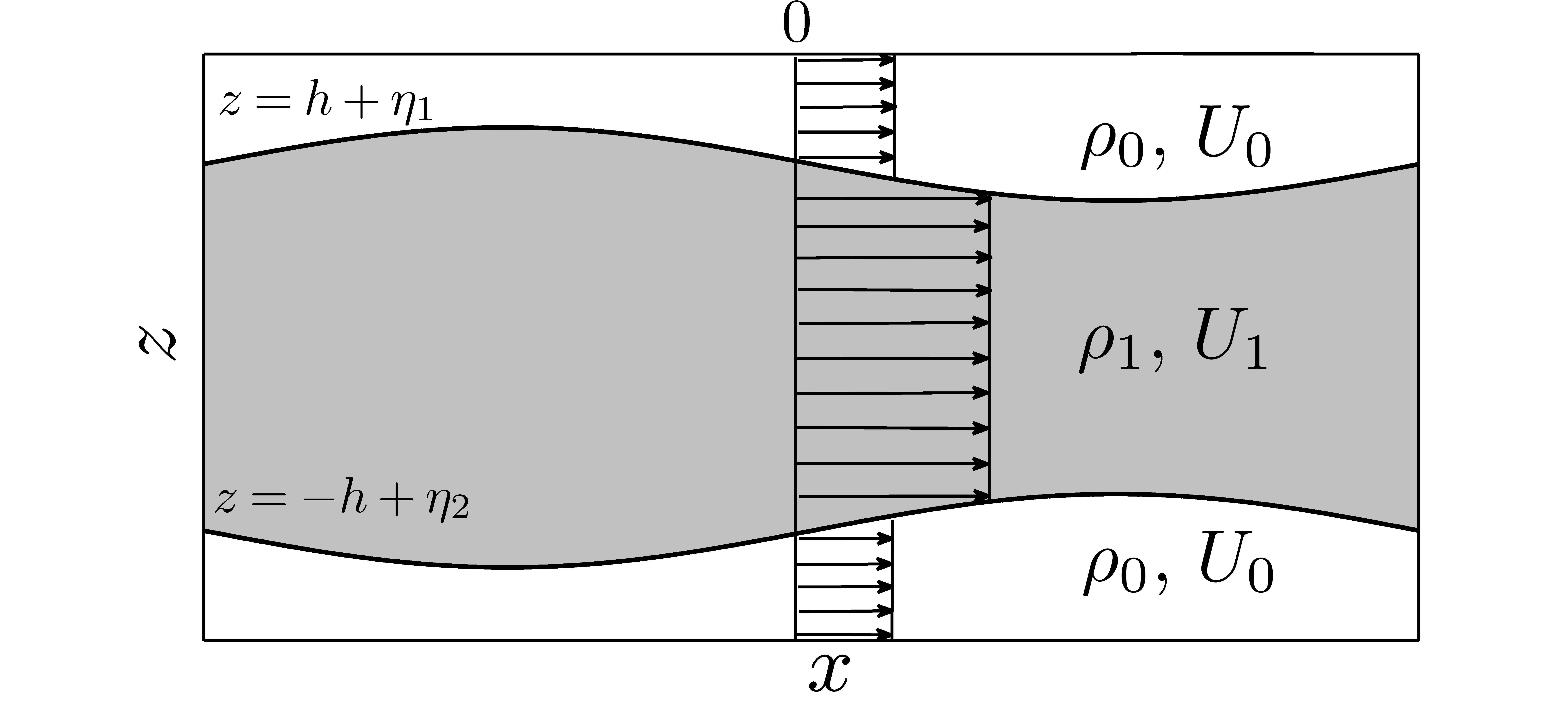}
\captionof{figure}{A schematic showing the initial configuration of a jet flow (a varicose initial condition has been chosen). Colors depict fluids of different densities. The flow is occurring in a horizontal plane ($g=0$).} 
\label{jet_schematic}
\end{figure}

Fluid flowing in the form of a jet stream is encountered in both natural flows like polar jet stream, as well as in industrial flows like fan spray nozzles, fuel injectors in diesel engines, etc. The system consists of a fluid of a given density flowing through a fluid of a different density; see figure \ref{jet_schematic}. There are density and velocity jumps across the two interfaces, which causes the latter  to roll up like  KH instability. Surface tension effects would be important for jets emanating from nozzles, but in this study those effects have been neglected. 
Presence of two interfaces leads to two modes of instabilities, viz. sinuous ($0$ phase shift between interfaces) and varicose ($\pi$ shifted interfaces) mode. We first derive the linear theory of the system. Then we show numerical simulations for various density jumps across the interfaces, and discuss its  effect on the roll up of the interfaces.



 The base velocity and density  profiles are as follows: 
 \begin{equation}
\bar{U}(z) = \left\{
        \begin{array}{cc}
        U_0 & \quad h \leq z \\ 
        U_1 & \quad  -h \leq z < h \\ 
        U_0 & \quad z<-h
        \end{array}
    \right.
   \qquad\qquad \bar{\rho}(z) = \left\{
        \begin{array}{cc}
        \rho_0 & \quad h \leq z \\ 
        \rho_1 & \quad  -h\leq z< h\\ 
        \rho_0 & \quad z<-h.
        \end{array}
    \right.
\end{equation}
           
           
We emphasize here that there is no gravity in the system (flows occurring in the horizontal plane). 
\subsubsection{Initial conditions}

Like previous cases, we formulate the problem as a system of linear homogeneous equations $ \mathcal{A}\mathcal{X} = 0 $.
The $ \mathcal{A} $ matrix is given by

\[
   \mathcal{A}=
  \left[ {\begin{array}{cccccc}
   -\alpha e^{-\alpha h} & 0 & 0 & 0 & i(\omega -  U_{0}\alpha) & 0 \\
   0 & -\alpha e^{-\alpha h} & \alpha e^{\alpha h} & 0 & i( \omega -  U_{1}\alpha) & 0 \\
    0 & -\alpha e^{\alpha h} & \alpha e^{-\alpha h} & 0 & 0 & i( \omega -  U_{1}\alpha) \\
    0 & 0 & 0 & \alpha e^{-\alpha h} & 0 & i( \omega -  U_{0}\alpha) \\
    A_{51} & A_{52} & A_{53} & 0 & 0 & 0 \\
   0 & A_{62} & A_{63} & A_{64} & 0 & 0
  \end{array} } \right],
\]
where
\begin{align*}
& & A_{51} = i\rho_{0} e^{-\alpha h}(-\omega + U_{0}\alpha )
, A_{52} = -i\rho_{1}e^{-\alpha h}(-\omega + U_{1}\alpha), \\
& & A_{53} = -i\rho_{1} e^{\alpha h}(-\omega + U_{1}\alpha )
 , A_{62} = i\rho_{1} e^{\alpha h}(-\omega + U_{1}\alpha ),\\
& & A_{63} = i\rho_{1} e^{-\alpha h}(-\omega + U_{1}\alpha)
 ,A_{64} = -i\rho_{0} e^{-\alpha h}(-\omega + U_{0}\alpha).
\end{align*}
The variable vector $ \mathcal{X} $ is given by
\[
\mathcal{X} = 
\begin{bmatrix}
C_{1} & C_{2} & C_{3} & C_{4} & \tilde{\eta_{1}} & \tilde{\eta_{2}}
\end{bmatrix}^{\dagger}.
\]
Similar to previous studies, we look for complex $ \omega $ and the corresponding $\mathcal{X}$ for obtaining the eigenstructure. There are two possible initial configurations - sinuous mode and the varicose mode.
The vortex strength of the interfaces are given by (\ref{eq:gamma1})-(\ref{eq:gamma2}).

\subsubsection{Results}

\begin{figure}
\centering
\includegraphics[width=1.0\linewidth]{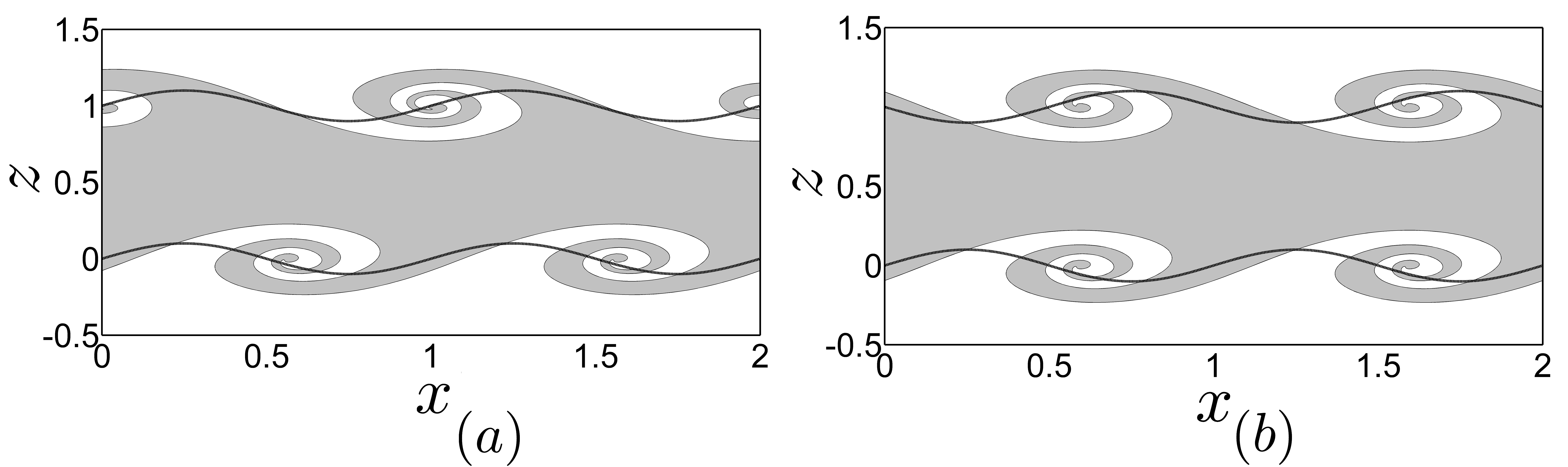}
\captionof{figure}{Nonlinear structures formed by a jet stream  using vortex method. (a) Initial phase difference of $ 0 $ (sinuous disturbance) and  (b) initial phase difference of $ \pi $ (varicose disturbance). Black curves represent the initial waves (initial amplitude has been exaggerated), while the gray region denotes the final structure.}
\label{fig:small}
\end{figure}

The nonlinear structures for a jet stream flowing in a stationary media or even in a media having a different velocity than the stream look similar to the KH roll ups. From linear theory we get two possible unstable modes - the sinuous mode and the varicose mode. If we start with one of these modes, the system starts to grow exponentially and yields nonlinear structure.  Figure \ref{fig:small} shows nonlinear structures corresponding to each of  these modes. The various parameters used for these simulations are: $ \rho_0/\rho_1 = 0.9 $,  $ U_0 = 0 $ $\mathrm{ms}^{-1}$, $ U_1 = 1 $ $\mathrm{ms}^{-1}$, $ \alpha=2\pi $ $\mathrm{m}^{-1}$, and  $ h = 0.5 $ m. Our results are very similar to the Direct numerical simulation results of \citet{minion1997performance} and \citet{hashimoto2013simulation}.  



\begin{figure}
\centering
\includegraphics[width=0.6\linewidth]{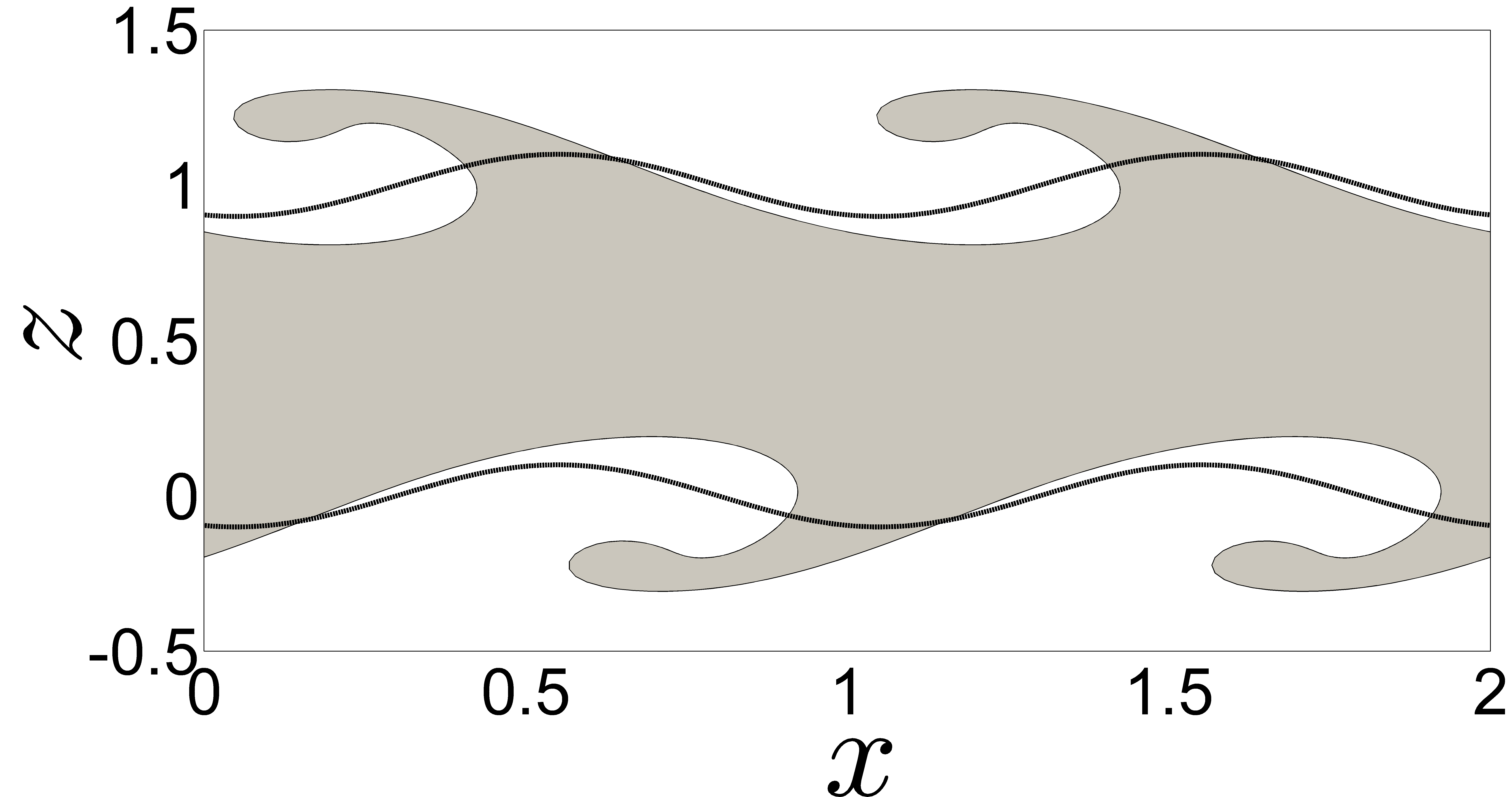}
\captionof{figure}{Nonlinear structures formed by an initially sinuous jet stream for $ \rho_0/\rho_1= 0.33 $.}
\label{jet_large_at}
\end{figure}
On increasing the density ratio, the interfaces show reduced tendency for KH like roll-ups; see  figure \ref{jet_large_at}. In this case $ \rho_0/\rho_1= 0.33 $, the rest of the parameters are same as before.




\section{Wave breaking in absence of shear: wave-topography interactions}
\label{sec:No_shear}

\subsection{Bragg resonance}
\begin{figure}
\centering
\includegraphics[width=0.8\linewidth]{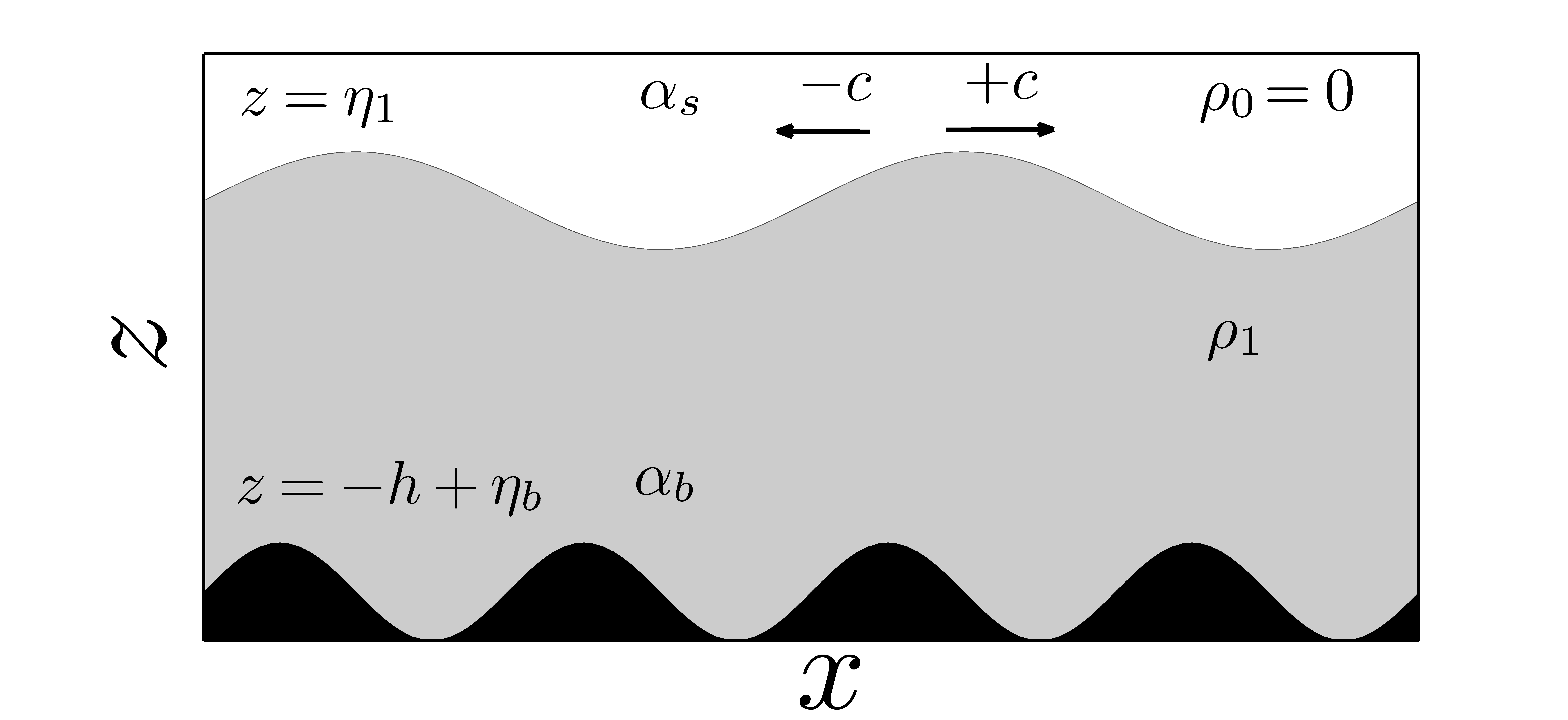}
\captionof{figure}{A schematic for the case of single layered  Bragg resonance. White region denotes air, gray denotes the fluid, and the black region denotes undulated bottom topography.}
\label{bragg_schematic}
\end{figure}

\begin{figure}
\centering
\includegraphics[width=0.8\linewidth]{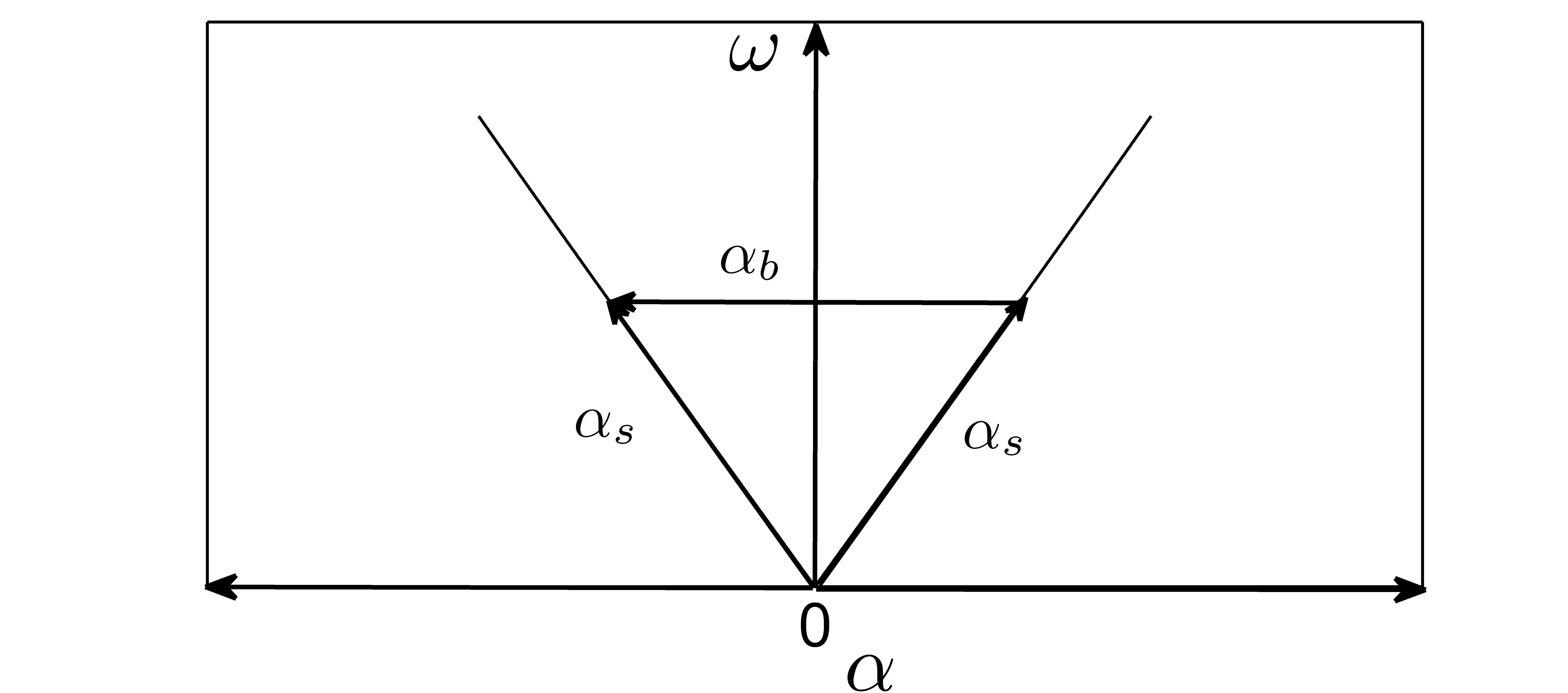}
\captionof{figure}{A Ball's diagram showing the condition for a single layer Bragg resonance.}
\label{bragg_ball}
\end{figure}

In resonant triad wave interactions, energy exchange occurs between three nonlinear waves. These resonances occur when some special conditions are satisfied - the wavenumber vectors as well as the corresponding wave frequencies add up to zero \citep{ball1964energy,craik1988wave}. Bragg resonance is a special case of such resonant triad interactions; it occurs when one of the three waves forming the resonant triad is a stationary undulated bottom topography (implying it is a wave of zero frequency). The undulated bottom acts as a mediator for the exchange of energy amongst the other two waves present in the system. The specific case considered here is a single layer fluid above a rippled bottom topography. Here  Bragg resonance can occur  between the two surface gravity waves (which are oppositely propagating with a speed of magnitude $c=\sqrt{g  \tanh(\alpha_s h)/\alpha_s}$, where $h$ is the  mean depth of the fluid and $\alpha_s$ is the wavenumber) present at the air-water interface and the
  bottom topography, as shown in figure \ref{bragg_schematic}. If we initialize with the configuration that a surface wave of wavenumber $ \alpha_{s} $ is propagating to the right over a rippled bottom of wavenumber $ \alpha_{b} $, then another wave of wavenumber $ \alpha_{s} $ traveling towards left would be generated. This resonance occurs under the condition 
\begin{equation}
\alpha_{b}=2\alpha_{s}.
\end{equation}
In figure \ref{bragg_schematic}, we have a surface wave of elevation $ \eta_{1} $ and wavenumber $ \alpha_{s} $, and a rippled bottom with elevation $ \eta_{b} $ and wavenumber $ \alpha_{b} $. The condition for a single layer Bragg resonance has been shown in figure \ref{bragg_ball}. Here  individual waves are  represented by  vectors in the $ \alpha-\omega $ plane. This representation is known as the Ball's diagram \citep{ball1964energy}.  

\subsubsection{Initial conditions}
The simulation of Bragg resonance using vortex method requires the initial interface elevation and vortex sheet strength. The initial conditions are exactly the same as the plunging breaker problem presented in section \ref{subsub:plung_init}, since we initialize the system with a shallow water surface gravity wave which interacts with the rippled bottom topography to generate an oppositely propagating surface gravity wave. The matrix $\mathcal{A}$ and the variable vector $\mathcal{X}$ are therefore the same as given in section \ref{subsub:plung_init}. Finally, the vortex strengths of the surface wave and the bottom are again given by  (\ref{eq:vortex_plunging1}) and (\ref{eq:vortex_plunging2}). 
\begin{figure}
\centering
\includegraphics[width=0.7\linewidth]{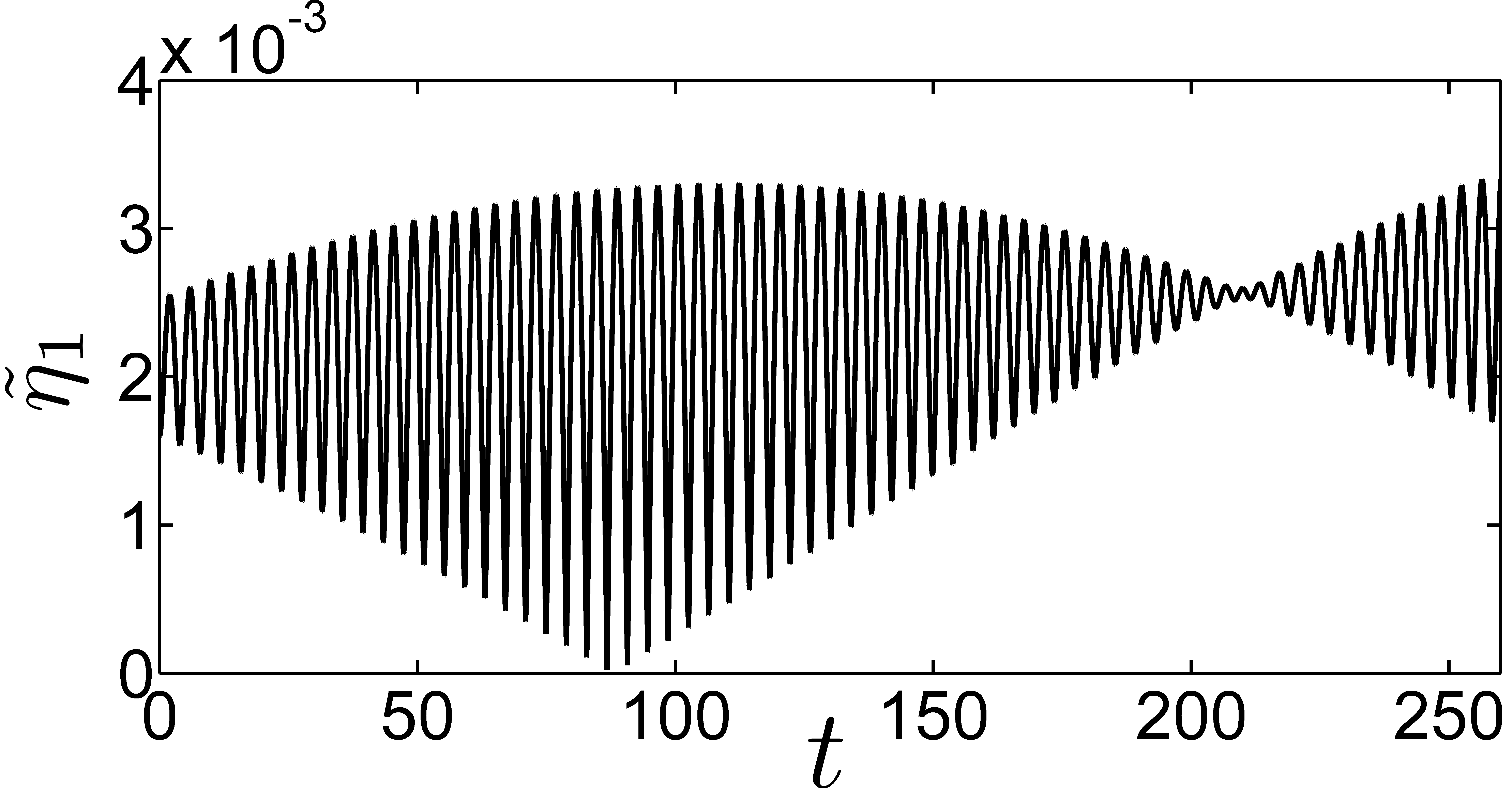}
\captionof{figure}{Amplitude $ \tilde{\eta}_1 $ (m) versus time $ t $ (s) corresponding to the surface wave.}
\label{bragg_amplitude}
\end{figure}

\subsubsection{Results}
The values of various parameters involved in the simulation are:  $ \alpha_{b}=4 $ $ \mathrm{m}^{-1} $, $ \alpha_{s}=2 $ $ \mathrm{m}^{-1} $, $\rho_0=0$ $\mathrm{kgm}^{-3}$, $ \rho_1 = 0.7877 $ $\mathrm{kgm}^{-3}$ ,  $ g = 1 $ $ \mathrm{ms}^{-2} $,  $ h=0.1 $ $ \mathrm{m} $ and initial amplitude of the surface wave $ \tilde{\eta}_{1}^{(0)} = 0.0016 $ $ \mathrm{m} $.

 Initially, there is only a rightward traveling wave of wavenumber $ \alpha_{s} $ present at the free surface. Resonant interaction with the rippled bottom gives rise to an oppositely traveling surface wave having the same wavenumber. The amplitude of the generated wave is initially zero and slowly increases with time, as shown in figure \ref{bragg_amplitude}. Since the frequency of both the original surface wave and the generated wave are same, they show rapid constructive and destructive interference to produce an amplitude variation similar to that observed for a standing wave. For a standing wave, the amplitude fluctuates rapidly between maximum value (constructive interference) and zero (destructive interference), and the envelope of the amplitude fluctuation stays constant. However in this case, the envelope of amplitude fluctuation  increases and then decreases because the amplitudes of the two oppositely propagating waves are unequal. It can be seen that around $ t=100 $ $ \mathrm{s} $ the envelope is maximum (varies from $ 0 $ to $ 2\tilde{\eta}_{1}^{(0)} $). Hence at this instant the amplitudes of the two waves are equal. After this, the amplitude of the generated wave becomes larger than the original surface wave, and the difference continues to increase until the amplitude of the original surface wave becomes zero and then the cycle repeats. 

\section{Summary and conclusions}
\label{sec:summary}

Various problems involving waves and hydrodynamic instabilities are conventionally studied in the linear regime using piecewise base velocity and density profiles. The advantage of using simplified profiles is that, on many occasions, they provide reasonably accurate qualitative and quantitative predictions of instabilities (as well as stable wave fields) in the initial stages. In fact, one can write analytical expressions for the range of instability, most unstable eigenvalue, corresponding eigenfunction, etc.\  in terms of the physical parameters of the problem \cite{drazin2004}. While this strategy has been very useful, it has largely remained confined to linear waves and instabilities (initial stages of the instability development). Previous works on extension of simplified profiles to the nonlinear regime were limited in the sense that they were applicable to the following cases: (i) homogeneous shear flows (contour dynamics technique)\cite{pozrikidis1985nonlinear}, (ii) infinite background shear (jump in background velocity) with or without density jump\cite{rosenhead1931formation,sohn2010long}, or (iii) density jump without any background velocity\cite{baker1982generalized,tryggvason1988numerical}.
In density stratified shear layers, it is quite common to find one or more density interfaces embedded in a background shear flow. 
The aim of this paper has been to find a strategy that can model such flows. We have shown that the classical vortex method under suitable modifications can capture the fully nonlinear structures emanating from stratified shear flows. The main mathematical difficulty in this case was  whether one can write unsteady Bernoulli's equation in the presence of background shear. We have formulated a ``shear-modified'' Bernoulli's equation, see (\ref{eq:top1})-(\ref{eq:bottom1}), which is the corner stone in deriving the vortex sheet evolution equation (\ref{eq:evol}). To the best of our knowledge, (\ref{eq:evol}) is the most general vortex sheet evolution equation known to us. Once the evolution of a vortex sheet embedded in background shear has been known, the next objective, in a mathematical sense, was to find the interaction equation between multiple vortex sheets (i.e.\ interfaces). In this regard we have derived Birkhoff-Rott equation for multiple interfaces, see (\ref{eq:birkhoff_disc_u})-(\ref{eq:birkhoff_disc_w}).

The mathematical model has been numerically implemented  following the techniques  outlined in \citet{sohn2010long}. In this numerical technique, known as vortex method, interfaces are represented by a discrete array of point vortices that interact with each other.  We have explored a wide variety of nonlinear phenomena - spilling (involves capillary-gravity waves) and plunging (involves long surface gravity waves) breakers, two classic stratified shear instabilities (Holmboe and Taylor-Caulfield), jet flows, as well as Bragg resonance (interaction of surface/interfacial gravity waves with rippled bottom topography, the latter being treated as a stationary wave). We observe that simple piecewise velocity and density profiles, when extended to the nonlinear regime, capture the essential nonlinear features, e.g.\ cusp formation and roll-ups observed in wave breaking, observed in experimental and/or involved numerical simulations with smooth, realistic profiles. 

Finally, we concentrate on the usefulness and applications of this  work. Since vortex method provides an exact (or near exact) solution of the (stratified) Euler equations, it is able to capture the correct nonlinear wave-mean feedback. This is of prime importance in the understanding of transition to turbulence in shear layers. Direct numerical simulations (DNS) are limited in this sense since they are only applicable to continuous profiles (no discrete interfaces allowed) and modest Reynolds numbers. Geophysical flows occur at very high Reynolds numbers, DNS of which would be extremely computationally expensive, perhaps impossible. Hence the ``shear-modified'' vortex method outlined in this paper would be useful, not only in capturing the detailed structures of high Reynolds number stratified shear flows, but also in understanding the mechanistic picture of wave-wave and wave-mean interactions. We emphasize here that there is a growing body of literature that tries to  pinpoint the mechanisms behind stratified shear instabilities (e.g.\ works of Heifetz and co-workers, see \citet{heif1999,heif2005,heifetz2009canonical,heifetzstratified2015}). This is because shear instabilities are often elusive and non-intuitive. Using piecewise profiles, these instabilities can be understood in a simplified manner in terms of resonant interaction of interfacial waves. While this wave interaction interpretation has been able to provide deep insights into shear instability mechanism, it has however remained limited to the linear regime. Our technique can very well extend the wave interaction interpretation to fully nonlinear regime, and can therefore help in providing a mechanistic understanding of the transition to turbulence.  
\bibliography{aipsamp}

\end{document}